\documentclass[letterpaper,UKenglish,numberwithinsect]{lipics-v2018-modified} \usepackage[full]{optional}

\opt{sub,final}{\renewcommand{\baselinestretch}{0.96}}

\usepackage
{hyperref}

\usepackage[textsize=scriptsize]{todonotes} 
\opt{final}{\setlength{\marginparwidth}{0.6in}} 
\opt{full}{\setlength{\marginparwidth}{1.5in}}

\usepackage{stackrel}

\usepackage[normalem]{ulem}

\usepackage{amsmath}
\usepackage{amssymb}
\usepackage{graphicx}

\usepackage[nomessages]{fp}

\usepackage{algorithm}
\usepackage[noend]{algpseudocode}

\usepackage{float}

\usepackage{comment}

\usepackage{color}
\usepackage{gensymb}

\usepackage{pgfplots} 

\makeatletter
\def\blfootnote{\xdef\@thefnmark{}\@footnotetext}
\def\BState{\State\hskip-\ALG@thistlm}

\makeatother

\newcommand{\removed}[1]{}

\newcommand{\vi}{\vec{i}}
\newcommand{\vc}{\vec{c}}

\usepackage{enumitem}

\renewcommand{\Pr}[1]{\mathrm{Pr}\left[#1\right]}
\newcommand{\E}[1]{\mathrm{E}\left[#1\right]}

\definecolor{DarkRed}{RGB}{182,11,1}

\usepackage{mathtools}

\newcommand{\ceil}[1]{\ensuremath{\left\lceil #1 \right\rceil}}
\newcommand{\floor}[1]{\ensuremath{\left\lfloor #1 \right\rfloor}}

\algnewcommand{\IfThenElse}[3]{
	\State \algorithmicif\ #1\ \algorithmicthen\ #2\ \algorithmicelse\ #3}

\def\logSizeErrorAagents{5.7}
\def\logSizeError{4.7}

\def\numStatesExponent{4} 
\def\timeExponent{2}
\def\constantNumGrv{5}
\def\cte{95} 

\renewcommand{\exp}[1]{\mathrm{exp}\left(#1\right)}

\newcommand{\polylog}{\mathrm{polylog}}
\newcommand{\N}{\mathbb{N}}

\newcommand{\ke}{\ensuremath{{k_\mathrm{ex}}}}

\def\R{\mathbb{R}}

\def\terminated{\texttt{terminated}}

\def\Cconstant{6}
\FPeval{\LCconstant}{clip(\Cconstant*2)}
\FPeval{\LCsubprotocolStateCountConstant}{clip(\Cconstant+\LCconstant)}
\FPeval{\Mconstant}{clip(\Cconstant*3)}
\newcommand{\aveconstant}{\Mconstant}
\newcommand{\countconstant}{\Cconstant}

\FPeval{\stateCountExponent}{clip(\Cconstant+\LCconstant+\Mconstant+\aveconstant+\countconstant)}

\def\CconstantMin{3}
\FPeval{\LCconstantMin}{clip(\CconstantMin*2)}
\FPeval{\LCsubprotocolStateCountConstantMin}{clip(\CconstantMin+\LCconstantMin)}
\FPeval{\MconstantMin}{clip(\CconstantMin*3)}
\newcommand{\aveconstantMin}{\MconstantMin}
\newcommand{\countconstantMin}{\CconstantMin}

\FPeval{\stateCountExponentMin}{clip(\CconstantMin+\LCconstantMin+\MconstantMin+\aveconstantMin+\countconstantMin)}

\def\exactCountingConvergenceTimeConstant{6}
\FPeval{\expectedTimeConstant}{clip(\exactCountingConvergenceTimeConstant+1)}

\newcommand{\phase}{\texttt{phase}}
\newcommand{\outpt}{\texttt{output}}

\def\ksum{\texttt{sum}}

\def\clock{\texttt{logSize2}} 
\def\doneC{\texttt{logSize2Generated}}
\def\x{\texttt{X}}
\def\a{\texttt{A}}
\def\f{\texttt{F}}
\def\s{\texttt{S}}
\def\role{\texttt{role}}
\def\passedGR{\texttt{updatedSUM}}
\def\doneG{\texttt{grGenerated}}
\def\agent{\texttt{agent}}
\def\doneEST{\texttt{protocolDone}}
\def\counter{\texttt{time}}
\def\gr{\texttt{gr}}

\def\phase{\texttt{epoch}}

\FPeval{\maxPhaseValue}{1184}
\FPeval{\betaUValue}{clip(12*\maxPhaseValue)}

\newcommand{\sizest}{\textsc{Log-Size-Estimation}}
\newcommand{\leaderless}{\textsc{Uniform-Leaderless-Phase-Clock}}
\newcommand{\partition}{\textsc{Partition-Into-A/F}}
\newcommand{\partitiontoAnS}{\textsc{Partition-Into-A/S}}

\newcommand{\epidemicClk}{\textsc{Propagate-Max-Clock-Value}}
\newcommand{\epidemicGr}{\textsc{Propagate-Max-G.R.V.}}
\newcommand{\timerDone}{\textsc{Check-if-Timer-Done-and-Increment-Epoch}}
\newcommand{\updateSum}{\textsc{Update-Sum}}
\newcommand{\movetonextGR}{\textsc{Move-to-Next-G.R.V}}

\newcommand{\epidemicPhase}{\textsc{Propagate-Incremented-Epoch}}

\newcommand{\genClk}{\textsc{Generate-Clock}}
\newcommand{\genGr}{\textsc{Generate-G.R.V}}

\newcommand{\reset}{\textsc{Restart}}

\newcommand{\false}{\textsc{False}}
\newcommand{\true}{\textsc{True}}
\newcommand{\agentA}{\text{agent-A}}

\newcommand{\rec}{\ensuremath{\textrm{rec}}}
\newcommand{\sen}{\ensuremath{\textrm{sen}}}

\algnewcommand{\LeftComment}[1]{\Statex \(\triangleright\) #1}

\newcommand{\calE}{\mathcal{E}}

\def\longrightharpoonup{\relbar\joinrel\rightharpoonup}
\def\longleftharpoondown{\leftharpoondown\joinrel\relbar}
\def\longrightleftharpoons{\mathop{\vcenter{\hbox{\ooalign{\raise1pt\hbox{$\longrightharpoonup\joinrel$}\crcr\lower1pt\hbox{$\longleftharpoondown\joinrel$}}}}}}
\def\rxn{\mathop{\rightarrow}\limits}  

\newcommand{\restateableLemma}[4]{
    \begin{lemma} \label{#1}
        #3
    \end{lemma}
    \expandafter\newcommand\csname restate#2\endcsname{
        \noindent{\bf Lemma~\ref{#1}.}
        \emph{#3}
    }
    \expandafter\newcommand\csname proof#2\endcsname{
        \begin{proof}
            #4
        \end{proof}
    }
}

\newcommand{\restateableTheorem}[4]{
    \begin{theorem} \label{#1}
        #3
    \end{theorem}
    \expandafter\newcommand\csname restate#2\endcsname{
        \noindent{\bf Theorem~\ref{#1}.}
        \emph{#3}
    }
    \expandafter\newcommand\csname proof#2\endcsname{
        \begin{proof}
            #4
        \end{proof}
    }
}

\newcommand{\restateableObservation}[4]{
    \begin{observation} \label{#1}
        #3
    \end{observation}
    \expandafter\newcommand\csname restate#2\endcsname{
        \noindent{\bf Observation~\ref{#1}.}
        \emph{#3}
    }
    \expandafter\newcommand\csname proof#2\endcsname{
        \begin{proof}
            #4
        \end{proof}
    }
}

\newcommand{\restateableCorollary}[4]{
    \begin{corollary} \label{#1}
        #3
    \end{corollary}
    \expandafter\newcommand\csname restate#2\endcsname{
        \noindent{\bf Corollary~\ref{#1}.}
        \emph{#3}
    }
    \expandafter\newcommand\csname proof#2\endcsname{
        \begin{proof}
            #4
        \end{proof}
    }
}

\title{Efficient size estimation and impossibility of termination in uniform dense population protocols}

\date{}

\opt{full,final}{
	\author
	{David Doty}
	{Department of Computer Science, University of California, Davis}
	{doty@ucdavis.edu}
	{}
	{\opt{full}{Supported by NSF grants 1619343 and 1844976.}}
    
    \author
    {Mahsa Eftekhari}
    {Department of Computer Science, University of California, Davis}
    {mhseftekhari@ucdavis.edu}
    {}
    {\opt{full}{Supported by NSF grants 1619343 and 1844976.}}
}

\opt{full}{
    
    \authorrunning{D. Doty and M. Eftekhari}
    
    \Copyright{David Doty and Mahsa Eftekhari}
    
    \opt{sub}{
        \subjclass{
            CCS 
            $\rightarrow$ 
            Theory of computation 
            $\rightarrow$ 
            Design and analysis of algorithms 
            $\rightarrow$ 
            Distributed algorithms
        }
    }
    \opt{full}{\subjclass{}}
    
    \opt{sub}{}
    \opt{full}{\keywords{}}
    
    \opt{sub}{
        \funding{
        DD, ME: NSF grant CCF-1619343.}
    }
    
    \opt{sub}{
        \EventEditors{Ulrich Schmid}
        \EventNoEds{1}
        \EventLongTitle{International Symposium on Distributed Computing}
        \EventShortTitle{DISC 2018}
        \EventAcronym{DISC}
        \EventYear{2018}
        \EventDate{Oct. 15 - Oct. 18, 2018}
        \EventLocation{New Orleans, LA}
        \EventLogo{}
        \SeriesVolume{1}
        \ArticleNo{1}
    }
    \opt{full}{
        \EventEditors{}
        \EventNoEds{0}
        \EventLongTitle{}
        \EventShortTitle{}
        \EventAcronym{}
        \EventYear{}
        \EventDate{}
        \EventLocation{}
        \EventLogo{}
        \SeriesVolume{}
        \ArticleNo{}
        \nolinenumbers 
    }
    
}

\begin{document}

\maketitle

\opt{sub,final}{\vspace{-0.4cm}}
\begin{abstract}
We study \emph{uniform} population protocols:
networks of anonymous agents whose pairwise interactions are chosen at random,
where each agent uses an \emph{identical} transition algorithm that does not depend on the population size $n$.
Many existing polylog$(n)$ time protocols for leader election and majority computation are nonuniform: 
to operate correctly, 
they require all agents to be initialized with an approximate estimate of $n$
(specifically, the value $\floor{\log n}$).

Our first main result is a uniform protocol for calculating 
$\log(n) \pm O(1)$ with high probability in 
$O(\log^\timeExponent n)$ time and 
$O(\log^\numStatesExponent n)$ states
($O(\log \log n)$ bits of memory).
The protocol is not \emph{terminating}:
it does not signal when the estimate is close to the true value of $\log n$.
If it could be made terminating with high probability,
this would allow composition with protocols 
requiring a size estimate initially. 
We do show how our main protocol can be indirectly composed with others in a simple and elegant way, 
based on \emph{leaderless phase clocks},
demonstrating that those protocols can in fact be made uniform.

However,
our second main result implies that the protocol \emph{cannot} be made terminating, 
a consequence of a much stronger result:
a uniform protocol 
for \emph{any} task requiring more than constant time 
cannot be terminating even with probability bounded above 0,
if infinitely many initial configurations are \emph{dense}: 
any state present initially occupies $\Omega(n)$ agents. 
(In particular no leader is allowed.)
Crucially, the result holds no matter the memory or time permitted.

\opt{full}{
    Finally, we show that \emph{with} an initial leader,
    our size-estimation protocol can be made terminating with high probability,
    with the same asymptotic time and space bounds.
}
\end{abstract}

\opt{sub}{\vspace{-0.8cm}}
\section{Introduction}
\label{sec:intro}


\emph{Population protocols}~\cite{AADFP06}
are networks that consist of computational entities called \emph{agents}
with no control over the schedule of interactions with other agents.
In a population of $n$ agents,
repeatedly a random pair of agents is chosen to interact,
each observing the state of the other agent before updating its own state.\opt{full}{\footnote{Using message-passing terminology, each agent sends its entire state of memory as the message.}}
They are an appropriate model for electronic computing scenarios such as sensor networks
and for ``fast-mixing'' physical systems such as
animal populations~\cite{Volterra26},
gene regulatory networks~\cite{bower2004computational},
and chemical reactions~\cite{SolCooWinBru08},
the latter increasingly regarded as an implementable ``programming language'' for molecular engineering,
due to recent experimental breakthroughs in DNA nanotechnology~\cite{chen2013programmable, srinivas2017enzyme}.


All problems computable with zero error probability 
by a constant-state population protocol are computable in $O(n)$ time~\cite{AAE08, DotHajLDCRNNaCo};
the benchmark for ``efficient'' computation is thus sublinear time,
ideally $\polylog(n)$.
For example, the transition $x,q \to y,y$ 
(starting with at least as many $q$ as the ``input'' state $x$)
computes $f(x) = 2x$ in expected time $O(\log n)$,
whereas $x,x \to y,q$
computes $f(x) = \floor{x/2}$
exponentially slower:
expected time $O(n)$~\cite{CheDotSolNaCo}.

Although the original model~\cite{AADFP06}
assumed a set of states and transitions that is constant with respect to $n$,
for important distributed computing problems such as
leader election~\cite{LeaderElectionDIST},
majority computation~\cite{alistarh2017time},
and computation of other functions and predicates~\cite{belleville2016time}
no constant-state protocol can stabilize in sublinear time with probability 1.\footnote{
    A protocol \emph{stabilizes} when it becomes unable to change the output.
    A protocol \emph{converges} in a given execution when the output stops changing,
    though it could take longer to subsequently stabilize.
    Known time lower bounds~\cite{LeaderElectionDIST, alistarh2017time, belleville2016time} are on stabilization, 
    not convergence.
    Recently Kosowski and Uznanski~\cite{kosowski2018population} achieved a breakthrough result,
    showing $O(1)$-state protocols for leader election and all decision problems computable by population protocols (the \emph{semilinear} predicates), 
    converging with high probability in $\polylog(n)$ time,
    and for any $\epsilon > 0$,
    probability 1 protocols for the same problems converging in $O(n^\epsilon)$ expected time.
    The latter protocols require $\Omega(n)$ time to stabilize, 
    as would any constant-state protocol due to the cited time lower bounds.
}
This motivated the study of protocols in which the set of states and transitions grows with $n$
(essentially adding a non-constant \emph{memory} to each agent).
Such protocols achieve leader election and majority computation using $O(\polylog(n))$ time,
keeping the number of states ``small'':
typically $O(\polylog(n))$~\cite{AG15, alistarh2017time, bilke2017brief, berenbrink2018simple, alistarh2018space},
although $O(\log \log n)$ states suffice for leader election~\cite{GS18}.

Unfortunately, many of these sublinear-time protocols~\cite{AG15, alistarh2017time, bilke2017brief, berenbrink2018simple, alistarh2018space}
are \emph{nonuniform}:
the set of states and transitions are allowed to depend arbitrarily on $n$
(this is not true of all, see for example recent fast, low-memory leader election protocols~\cite{GS18,gkasieniec2018almost}).
This capability is used to initialize each agent with an approximate estimate of $n$
(the value $\floor{\log n}$) required by the protocols.
\opt{full}{
    A representative example portion of such a protocol is shown in Fig~\ref{fig:uniform}:
    each agent has an internal ``counter'',
    which increments upon each encounter with an $x$.
    When the counter reaches $\log n$,
    the protocol terminates (or moves to a different ``stage'').
}

\opt{full}{
    \begin{figure*}[h]
      \centering
        \includegraphics[draft=false,width=\textwidth]{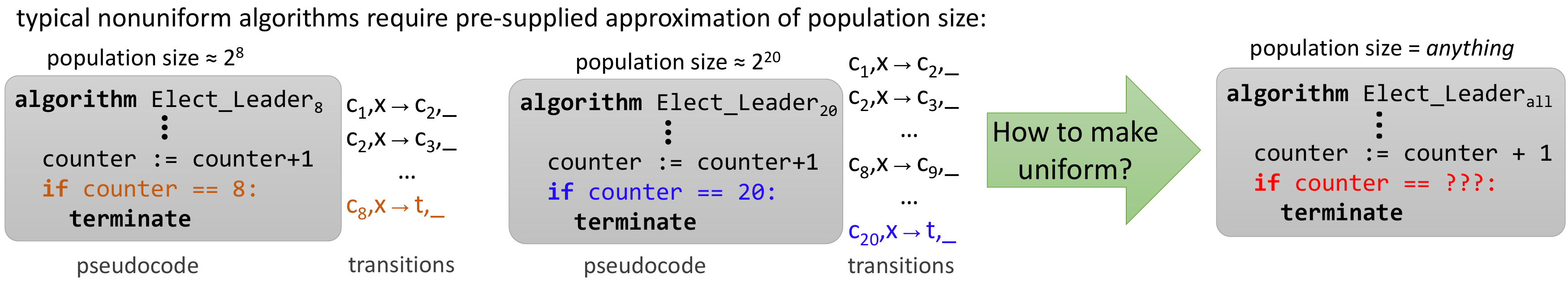}
        \caption{
          Many population protocols with $\omega(1)$ states use nonuniform algorithms:
          the value $\log_2 n$ is ``hardcoded'' into the reactions.
          Above, ``{\tt terminate}'' could mean ``end the whole algorithm'',
          ``move to the next stage'',
          or simply
          ``stop increasing {\tt counter}''.
        }
        \label{fig:uniform}
    \end{figure*}
}

More desirable would be a \emph{uniform} protocol
in which each agent's local algorithm for computing the outputs, 
given the inputs, has no knowledge of $n$.
Such an algorithm may produce outputs longer than its inputs,
retaining the ability to use a number of states that grows with the population size.
A uniform protocol can be deployed into \emph{any} population without knowing in advance the size,
or even a rough estimate thereof.


\opt{sub,final}{\vspace{-0.2cm}}
\subsection{Contributions}



Nonuniform protocols in the literature~\cite{AG15, alistarh2017time, bilke2017brief, berenbrink2018simple, alistarh2018space} initialize each agent with the value $\floor{\log n}$.
Hence we study the problem of computing an approximate estimate of $\log n$.

Our first main result, 
Theorem~\ref{thm:main-size-estimation-protocol}, 
is a uniform protocol,
starting from a configuration where all $n$ agents are in an identical state,
that with high probability computes $\log n \pm O(1)$
(storing the value in every agent),
using $O(\log^\timeExponent n)$ time and $O(\log^\numStatesExponent n)$ states.\opt{full}{\footnote{
    It appears difficult to compute $\floor{\log n}$ exactly,
    rather than within a positive additive constant,
    since for all $k$,
    such a protocol could distinguish between the exact population sizes $2^k-1$ and $2^k$.
    Note also that our protocol has a positive probability of error.
}}
This answers affirmatively open question 5 of~\cite{doty2018exact}.
This is done primarily by generating a sequence of geometric random variables,\footnote{
    To our knowledge, this constitutes the first analysis of sums of independent random variables,
    each of which is a maximum of geometric random variables.
    Standard Chernoff and other tail bounds generally used for bounded random variables
    fail in this case.
    We apply the theory of sub-exponential random variables~\cite{rigollet2015highdimensional} 
    to obtain strong bounds on the moment-generating function of a maximum of geometric random variables in order to obtain the required Chernoff bounds.
}
and propagating the maximum to each agent.
However, before the maximum reaches all agents they begin computation;
thus we use a restart scheme similar to~\cite{GS18} to reset an agent's computation when it updates to a higher estimate of the max.

One might hope to use this protocol as a subroutine to
``uniformize'' existing nonuniform protocols for leader election and majority~\cite{AG15, alistarh2017time, bilke2017brief, berenbrink2018simple, alistarh2018space}.\footnote{
    Some protocols for leader election~\cite{GS18,gkasieniec2018almost} are uniform, 
    but other protocols~\cite{AG15, alistarh2017time, bilke2017brief, berenbrink2018simple} 
    have the benefit of simplicity 
    and may possibly be easier to reason about and compose with other protocols.
}
Suppose the size-estimating protocol could be made terminating,
eventually producing a termination ``signal'' that with high probability does not appear until the size estimate has converged.
This would allow composition with other protocols requiring the size estimate.
It has been known since the beginning of the population protocol model~\cite{AADFP06}
that termination cannot be guaranteed with probability 1.
However, leader-driven protocols can be made terminating with high probability,
including simulation of register machines~\cite{AAE08} or exact population size counting~\cite{M15}.

Our second main result,
Theorem~\ref{thm:main-termination},
shows that this is impossible to do with our leaderless size-estimation protocol
and a very wide range of others.
This answers negatively open questions 1-3 of~\cite{doty2018exact}.
The production of such a terminating signal cannot be delayed,
even with probability bounded above 0,
by more than $O(1)$ time in any uniform protocol where,
for some $\alpha>0$,
infinitely many valid initial configurations are \emph{$\alpha$-dense},
meaning that each state present is the state of at least $\alpha n$ agents.
This holds even for randomized protocols with a nondeterministic transition function.
(Because this is an impossibility result, the fact that it holds for both deterministic and randomized protocols makes it stronger than if it held only for deterministic protocols.)
Since virtually all non-trivial computation with population protocols requires $\Omega(\log n)$ time\footnote{
    $\Omega(\log n)$ is a lower bound on most interesting computation:
    by a coupon collector argument,
    this is the expected time for each agent to have at least one interaction.
}
(including leader election, and computation of predicates and functions such as majority and $g(x)=2x$),
this implies that no uniform terminating protocol can solve these problems from dense initial configurations.

The hypothesis of density is crucial:
with a \emph{leader},
high-probability termination is possible in a uniform protocol~\cite{AAE08}.
The hypothesis of uniformity is also crucial:
if each agent can \emph{initially} store a value $f(n)$, 
then a termination signal can be delayed until some agent experiences $f(n)$ interactions,
an event whose expected time grows unboundedly with $n$ if $f$ grows sufficiently fast.
This result uses a density argument similar to that used previously to show time lower bounds,
which assume a state set of size $O(1)$~\cite{Do14, LeaderElectionDIST, belleville2016time}
or $\leq \frac{1}{2} \log \log n$~\cite{alistarh2017time}.
In contrast, our argument holds for \emph{any} state set size,
by showing that a particular subset of states is produced in constant time w.h.p.,
and using a careful argument to show that this subset necessarily contains the termination signal.

Despite this difficulty in directly composing size estimation with a downstream protocol (or several stages/subprotocols composed in series),
we present a general and simple method of composition (via restarting), 
based on a ``leaderless phase clock'' using a weaker log population size estimate $s$ (called $\clock$ in the pseudocode in Section~\ref{subsec:protocol}) obtained initially 
(where $\log n - \log \ln n \leq s \leq 2 \log n$ w.h.p.).\footnote{
    The first leaderless phase clock for population protocols was proposed in~\cite{alistarh2018space}.
    Ours is different, based on~\cite{sudo2018logarithmic}.
    Both are nonuniform, relying on an estimate of $\log n$.
}
Based on $s$ and the expected convergence time of the downstream protocol, 
each agent once per interaction increments a counter $c$, 
from $0$ up to $f(s)$,
and the first agent to reach $f(s)$ signals the entire population to terminate (or move to the next stage).
$f(s)$ is chosen large enough that no agent reaches $f(s)$ before the downstream protocol converges.
The entire downstream protocol is reset if the initial size estimate $s$ changes.
With the above scheme, agents need to store the variables
$s$, $c$, and possibly also $f(s)$ 
(in our case $f(s) = O(s)$ so it need not be stored explicitly, but if $f(s) = \mathrm{poly}(s)$, for example, $f(s)$ may need to be stored separately from $s$).
If the downstream protocol requires $t(n)$ time to converge, then agents also set their threshold $f(s) > t(n)$ 
(where $f(s)$ is ``large'' compared to $t(n)$).
This requires $O(f(s)^2 \cdot \log n)$ states will be added to the state complexity of the protocol,
or $O(\log^2 n)$ if $f(s) = O(\log n)$ (as in our case) since $f(s)$ need not be stored explicitly.
To compose multiple downstream stages/subprotocols in series,
we also need a way to compute and possibly store the number $K$ of stages (in our case $K = \Theta(\log n)$, also chosen as a constant times $s$, so $K$ need not be stored explicitly),
and we need to store an index indicating which stage we are on.
For $K$ stages,
this multiplies the state complexity by $K$ if $K = O(\log n)$ and 
$K^2$ otherwise (since $K$ must be stored explicitly in the latter case).


\opt{sub,final}{\vspace{-0.2cm}}
\subsection{Related work}

The work of this paper was inspired by recent work on nonuniform polylog time leader election/majority~\cite{AG15, alistarh2015fastExactMajority, alistarh2017time, bilke2017brief, berenbrink2018simple, alistarh2018space, sudo2018logarithmic};
the fact that those protocols require an approximate size estimate is the direct motivation for seeking a protocol that can compute such an estimate
(though unfortunately due to Theorem~\ref{thm:main-termination},
composition of our protocol with these is not totally straightforward).

Some nonuniform protocols
crucially rely on an estimate of $\log n$ (e.g.~\cite{AG15, alistarh2017time, bilke2017brief, berenbrink2018simple, alistarh2018space, sudo2018logarithmic}) for \emph{correctness}. 
Other nonuniform protocols are more robust, 
using the estimate merely to allow the protocol to have a finite number of states. 
For example, Alistarh and Gelashvili~\cite{AG15} show a $O(\log^3 n)$-time protocol for leader election in which leaders increment a counter on each interaction. 
The uniform variant of that protocol, with no estimate of $\log n$, 
is correct with probability 1, 
and the estimate of $\log n$ is used only to bound the counter 
(hence also the number of states) below $\log n$. 
Nevertheless, it is not obvious how to modify that protocol to be uniform and have a bounded number of states with high probability.

\noindent
{\bf Self-stabilizing leader election and exact size counting.}
Cai, Izumi, and Wada~\cite{cai2009space}
(using different terminology)
show an impossibility result for uniform population protocols,
that no protocol 
electing a leader can be uniform
if it is also required to be
\emph{self-stabilizing}: 
correct with probability 1 from \emph{any} initial configuration.
In fact, it must be nonuniform in a very strong way:
the \emph{exact} population size must be encoded into each agent.
Self-stabilizing exact size computing has also been shown to be possible with a leader~\cite{beauquier2015space} in 
$O(n \log n)$ time and $O(n)$ states for the leader 
and $2$ states for the other agents, 
all asymptotically optimal parameters in the self-stabilizing setting~\cite{AspnesBBS2016}.

\noindent
{\bf Exact size counting.}
In the less restrictive setting where all agents start from a pre-determined state,
Michail \cite{M15} proposed a uniform \emph{terminating} protocol 
(where agents ``know'' when they have converged)
in which a pre-elected leader 
computes the exact population size $n$ in $O(n\log{n})$ time with high probability.
Going from the terminating to the less restrictive \emph{converging} criterion 
(where agents eventually converge on the correct size, 
but do not know when this occurs), 
exact size counting is possible in 
$O(\log n \log \log n)$ time and $O(n^{60})$ states~\cite{doty2018exact},
\emph{without} an initial leader.

\noindent
{\bf Approximate size estimation.}
Alistarh, Aspnes, Eisenstat, Gelashvili, and Rivest~\cite{alistarh2017time} show a
uniform protocol that in $O(\log n)$ expected time and states converges to an approximation $n'$ of the population size $n$,
computing an integer $k$ 
such that with high probability $\frac{1}{2} \log n \leq k \leq 9 \log n$,
i.e., $\sqrt{n} \leq 2^{k} \leq n^9$.
Each agent generates (an approximation of) a geometric random variable, letting $k$ be their maximum.
We use their protocol as the first step of ours.
The analysis of~\cite{alistarh2017time} is based on synthetic coins with a deterministic transition function, which have bias complicating the analysis. 
Our randomized model assumes access to perfectly random bits,
so a simpler analysis (Corollary~\ref{cor:grvalue}) shows that $\log n - \log \ln n \leq k \leq 2 \log n$ w.h.p.
The remainder of our protocol improves this from a constant multiplicative error in approximating $\log n$
to a constant \emph{additive} error.
In other words we estimate the population size to within a constant multiplicative factor
(instead of a polynomial factor as in~\cite{alistarh2017time}),
but use 
$O(\log^\timeExponent n)$ time and 
$O(\log^\numStatesExponent n)$ states.

Berenbrink, Kaaser, and Radzik~\cite{berenbrink2019oncounting}
independently studied the same size estimation problem as ours, 
obtaining stronger bounds on additive error and number of states:
computing the value $\floor{\log n}$ or $\ceil{\log n}$ (i.e., additive error $< 1$) with high probability, 
using $O(\log^2 n)$ time and $O(\log n \log \log n)$ states.
They also show a protocol with probability 1 of correctness, using $O(\log^2 n)$ time and $O(\log^2 n \log \log n)$ states.

\opt{sub}{\vspace{-0.5cm}}
\section{Model}
\label{sec:model}

To formally define \emph{uniform} computation in population protocols,
the agents' transition algorithm is modeled as a 2-tape deterministic Turing machine (TM) with the read only ``input tape'' as tape 1 
(for reading the other agent's state)
and read-write ``working tape'' as tape 2 
(for storing this agent's state).\footnote{
    Our model generalizes the original constant-state model~\cite{AAE06} by allowing the memory potentially to grow with $n$;
    however, constant-state protocols can be implemented with our model.
    It is worth distinguishing four ways for memory to increase with $n$:
    1) not at all 
    (constant-state),
    2) increasing with $n$ but, for each $n$, bounded by a constant depending on $n$ 
    (most non-uniform protocols),
    3) possibly unbounded but bounded with probability 1 
    (this paper),
    and
    4) unbounded with positive probability.
}

Our protocol describes a constant number of integer fields comprising each agent's state,
which could all be stored in the working tape and separated by a special symbol.
An agent's working tape is identical to what it was on the conclusion of the previous interaction.
When two agents interact, each copies the content of the other's tape 2 its own tape 1,
and then each of their TM states is reset from a halting TM state to the start TM state.
The space usage (in bits) $s$ is defined as normal for TMs:
the maximum number of tape cells that are written during the computation on the read/write working tape.
The number of possible agent states (working tape contents) is then $c^s$,
where $s$ is the maximum space usage of any agent during an execution of the protocol and $c$ is the size of the tape alphabet.
For ease of understanding,
we will use standard population protocol terminology and not refer explicitly to details of the TM definition except where needed.
A \emph{state} $s \in \Lambda$ always refers to the TM working tape content of an agent
(leaving out TM state and tape head positions since these are identical in all initial configurations),
where $\Lambda$ is the set of all agent states.
A \emph{configuration} $\vc \in \N^\Lambda$ is a vector indexed by a state,
where $\vc(s)$ is the \emph{count} of state $s$ in the population. 
We set the output of our protocol the value stored in a special field labeled ``output''. 
Some definitions allow the output to be a function of the fields stored in an agent's memory,
without the output itself counting against the memory usage.
Our protocol reuses a field for the output that is used prior in the protocol,
so our memory usage is the same under either definition.


We furthermore assume that each agent has access to independent uniformly random bits,
assumed to be pre-written on a special read-only tape
(this allows the TM to be deterministic even though it is computing a nondeterministic relation).
This is different from the traditional definition of population protocols,
which assumes a deterministic transition function.
In our case, 
we have a transition \emph{relation} $\delta \subseteq \Lambda^4$.
Several papers~\cite{alistarh2017time, berenbrink2018simple}
indicate how to use the randomness built into the interaction scheduler to provide nearly uniform random bits to the agents,
using various \emph{synthetic coin} techniques,
showing that the deterministic model can effectively simulate the randomized model.
In the interest of brevity and simplicity of presentation,
we will simply assume in the model that each agent has access to a source of uniformly random bits.
A variant of our protocol using the sender/receiver choice to simulate uniformly random bits with a deterministic transition function,
with the same time, state, and error bounds, is available 
\opt{final}{in~\cite{doty2018efficientArxiv}.}
\opt{full}{in Section~\ref{app:protocol-det}.}

Throughout this paper, $n$ denotes the number of agents in the population.
Repeatedly, a pair of agents is selected uniformly at random to interact,
where they run the transition algorithm on the pair of states they were in prior to the interaction,
and storing the output states for their next interactions.
The \emph{time} until some event is measured as the number of interactions until the event occurs, divided by $n$,
also known as parallel time.
This represents a natural model of time complexity in which we expect each agent to have $O(1)$ interactions per unit of time, hence across the whole population, $\Theta(n)$ total interactions occur per unit time.
All references to ``time'' in this paper refer to parallel time.
An \emph{execution} is a sequence of configurations $\vc_0,\vc_1,\ldots$
such that for all $i$,
applying a transition to $\vc_i$ results in $\vc_{i+1}$.
$\log n$ is the base-2 logarithm of $n$,
and $\ln n$ is the base-$e$ logarithm of $n$.

\subsection{Definition of correctness and time}
\label{sec:correct-defn-variations}

The notion that a protocol's configuration ``has the correct answer'' is problem-specific.
For leader election, it means there is a single leader agent.
For predicate computation, it means all agents have the correct Boolean output.
In this paper, since our goal is to approximate $\log n$ within additive factor $\logSizeErrorAagents$,
we say a configuration is \emph{correct} if the $\outpt$ field of each agent is within $\logSizeErrorAagents$ of $\log n$.\footnote{
    We note that our notion of function approximation differs from that of Belleville, Doty, and Soloveichik~\cite{belleville2016time}.
    They use a \emph{distributed} output convention, 
    where the output of a function $f:\N^d \to \N$ is encoded as the population count of agents in a special output state $y$.
    Thus one must examine the entire population to know the output.
    In our \emph{local} output convention,
    each agent has a field encoding a value from the function's range.
    The output is undefined if some agents have different values,
    and defined to be their common value otherwise.
    This is similar to how Boolean predicate output with range $\{0,1\}$ is encoded in population protocols~\cite{AADFP06}.
}

The following definitions match those used in the literature, when other notions of ``correct'' are substituted.
Let $\mathcal{E} = (\vc_0,\vc_1,\ldots)$ be an infinite execution.
A configuration $\vc$ is \emph{stably correct} if every configuration reachable from $\vc$ is correct.\footnote{
    Belleville, Doty, and Soloveichik~\cite{belleville2016time}
    also consider function approximation,
    but define a configuration to be stable if the output \emph{cannot} change,
    whereas we allow it to change within a small interval around the correct value.
    The time lower bound techniques of~\cite{belleville2016time} do not apply to our more relaxed notion of stability.
}
We say $\mathcal{E}$ \emph{converges} at interaction $i$ if $\vc_{i}$ is not correct and for all $j > i$, $\vc_j$ is correct.
We say $\mathcal{E}$ \emph{stabilizes} at interaction $i$ if $\vc_{i}$ is not stably correct and for all $j > i$, $\vc_j$ is stably correct.
A protocol can converge and/or stabilize with probability 1 or a smaller probability.
However, if the set of reachable configurations is bounded with probability 1 
(which is the case for the protocols discussed in this paper),
then for any $p \in [0,1]$, 
a protocol converges with probability $p$ if and only if it stabilizes with probability $p$.\footnote{
    Let $C$ and $S$ respectively be the set of stabilizing and converging executions.
    Clearly $S \subseteq C$.
    Although $S \subsetneq C$ is possible,
    we argue that $\Pr{C \setminus S} = 0$.
    Suppose a protocol converges in an execution $(\vc_0,\vc_1,\ldots)$ 
    at interaction $i$ (so $\vc_j$ is correct for all $j > i$).
    If did not stabilize,
    then for all $j > i$, some incorrect configuration $\vec{d}_j$ would be reachable from $\vc_j$.
    Let $p_j > 0$ denote the probability of reaching $\vec{d}_j$ from $\vc_j$.
    The set of reachable configurations is bounded with probability 1,
    so $\min\limits_{j > i} p_j$ is well-defined and positive.
    The probability of never reaching any $\vec{d}_j$ is then 0.
}
For a computational task $T$ equipped with some definition of ``correct'',
we say that a protocol $\mathcal{P}$ \emph{stably computes $T$ with probability $p$} if,
with probability $p$, it stabilizes (equivalently, converges).
If $p$ is omitted, it is assumed $p=1$.
However, when measuring time complexity, convergence and stabilization may be much different.
We say that $\mathcal{P}$ converges (respectively, stabilizes) in (parallel) time $t(n)$ with probability $p$ if, 
with probability $p$, it produces an execution that converges (resp., stabilizes) by interaction $i$, where $i/n \leq t(n)$. 
Many protocols converge much faster than they stabilize, 
such as those that combine a fast, error-prone subprotocol with a slow, error-free protocol, e.g.,~\cite{AAE08, kosowski2018population, CheDotSolNaCo}.
However, for the protocol of this paper, convergence and stabilization coincide.
We use the term ``converge'' throughout the paper to refer to this event.

Many papers separately measure high-probability time convergence and expected time to converge.
Our protocol has positive probability of error,
but we argue that expected time is a meaningful notion only with error probability 0,
which is why we do not measure expected time.
The only reasonable definition of ``time until correctness'' on a non-converging execution is $\infty$.
So with $\Pr{\text{doesn't converge}} > 0$, 
the expected convergence time is 
$
    \E{\text{time}|\text{converges}} 
    \cdot \Pr{\text{converges}} 
    +
    \E{\text{time}|\text{doesn't converge}}
    \cdot \Pr{\text{doesn't converge}}
= 
    \E{\text{time}|\text{converges}}
    \cdot \Pr{\text{converges}}
    + 
    \infty
=
    \infty.
$
One could imagine measuring only $\E{\text{time}|\text{converges}}$.
However, conditioning can artificially ``speed up'' the process.\footnote{
    Consider a hypothetical protocol that runs a parallel subprotocol $S$ that completes quickly and, upon completion, 
    somehow prevents the main protocol $M$ from converging.
    The main protocol, on the other hand, may somehow detect if it completes before $S$ does, and if so, $M$ then shuts $S$ down.
    Many executions will not converge,
    but those that do must be very fast in order to converge before $S$ completes.
    Thus conditioning on convergence ``anthropically speeds up'' convergence~\cite{aarsonson2006anthropic}.
    This is an extreme example that has the property that the probability of correctness is reduced by $S$, but it nevertheless shows that measuring conditional expected time can be problematic.
}

\section{Fast protocol for estimating $\log n$ within $O(1)$ additive error}
\label{sec:protocol}

In this section we describe a uniform protocol for computing the value of $\log n$ with an additive error,
i.e., estimating the population size to within a constant multiplicative factor.
We say a population protocol is \emph{leaderless} if all agents start in the same state.
\restateableTheorem{thm:main-size-estimation-protocol}{THMmainSizeEstimationProtocol}
{   
    There is a uniform leaderless population protocol that 
    converges in time $O(\log^\timeExponent n)$ 
    with probability $\geq 1-1/n^2$, 
    uses $O(\log^\numStatesExponent n)$ states
    with probability $\geq 1-{O(\log n)}/{n}$, 
    and
    stores in each agent an integer $k$ such that 
    $|k - \log n| \leq \logSizeErrorAagents$
    with probability $\geq 1-9/n$.
}{
    By Corollary~\ref{cor:timecomplexity}
    \sizest\ take $O(\log^2 n)$ time to converge with probability at least $1-1/n^2$.
    By Lemma~\ref{lem:space}, \sizest\ protocol uses $O(\log^\numStatesExponent n)$ states with probability at least $1-O(\log n)/n$.
    
    Finally, Lemma~\ref{lem:correctnessofProtocol} guaranties the \outpt\ obtained by \a\ agents is with an additive error $\logSizeErrorAagents$ of $\log n$ with probability at least $1-9/n$.
}


We note that the protocol has a positive probability of error.
It is open to find a protocol using $\polylog(n)$ time/states 
computing
$\log(n) \pm O(1)$
with probability 1.

The protocol is described and its time and state complexity analyzed in Subsection~\ref{subsec:protocol}.
Much of the analysis of the approximation involves proving a bound on the moment-generating function of a maximum of geometric random variables,
enabling the Chernoff technique can be applied to sums of such variables.
This is quite nontrivial; 
\opt{final}{see appendix in~\cite{doty2018efficientArxiv}}
\opt{full}{see Section~\ref{subsec:geom}}.

\subsection{Intuition}
Alistarh et al.~\cite{alistarh2017time} describe a protocol for estimating $\log n$ 
within a constant multiplicative factor. 
A $\frac{1}{2}$-geometric random variable is the number of flips needed to get one head when flipping a fair coin. 
In their protocol, each agent generates an independent geometric random variable $\mathbf{G}_i$,
then propagates the maximum 
$\mathbf{M} = \max_{1 \leq i \leq n} \mathbf{G}_i$
by \emph{epidemic}:
transitions of the form $i,j \to j,j$ for $i \leq j$, 
which in $O(\log n)$ time ``infect'' all agents with the maximum.
It is known that $\E{\mathbf{M}} \approx \log n$~\cite{eisenberg2008expectation},
and $\log n - \log \ln n \leq \mathbf{M} \leq 2 \log n$ with probability 
$\geq 1-O(1)/n$ (Lemma C.7~\cite{doty2018efficientArxiv}).


We take the obvious extension of this approach:
do this $K$ times and take an average.
The estimated average is within $O(1)$ of $\log n$ so long as $K = \Omega(\log n)$
\opt{final}{\cite{doty2018efficientArxiv}.}
\opt{full}{(Corollary~\ref{cor:sumofgeomConcreteK}).}
One problem to solve first is how to calculate $K$;
after all, $K = \Theta(\log n)$ scales with $n$,
so with a uniform protocol it cannot be encoded into the agents at the start.
The agents estimate it using the protocol of~\cite{alistarh2017time}.
Since that protocol is converging but not terminating
(provably it cannot be made terminating by Theorem~\ref{thm:main-termination}),
each time an agent updates its value of $K$,
it reinitializes the remainder of its state.

However, a trickier problem remains: 
a na\"{i}ve approach to implement ``averaging of $K$ numbers'' requires storing $K = \Theta(\log n)$ numbers in each agent,
each having value $\Theta(\log n)$,
implying the number of states is $\Theta((\log n)^{\log n}) = \Theta(n^{\log \log n})$.
This is even more than the $O(n^{60})$ sufficient to quickly compute \emph{exactly} $n$~\cite{doty2018exact}.
To overcome this problem, we use a ``leaderless phase clock'' similar to those of~\cite{alistarh2018space,sudo2018logarithmic,mocquard2018population}, but uniform.
Unlike the phase clock used 
by~\cite{alistarh2018space,mocquard2018population}, our  leaderless phase clock 
simply increments a counter on every interaction.
This simultaneously gives an elegant way to compose our protocols with downstream protocols requiring the size estimate.
Agents count their number of interactions and compare it with a threshold value $\Theta(\log n)$.
Whenever their number of interaction passes the threshold they will move to the next round similar to the protocol described above (the population with a leader). 
The threshold is calculated in the following way.
In our protocol, agents start generate a geometric random variable called \clock\ and propagate the maximum \clock\ among themselves. 
After agents agree on the $\clock$ variable, a constant multiple $\cte . \clock$ is the threshold in their leaderless phase clock.
This lets the agents synchronize epochs of the algorithm,
each taking $O(\log n)$ time,
and prevent the next epoch from starting until the previous has concluded.

The probabilistic clock inside agents might go off very soon at the very beginning of the protocol, 
but after $O(\log n)$ time all agents will store the maximum generated \clock\ and their leaderless phase clock will eventually converge to a stable one which goes off after completion of a predefined constant factor of $\log n$;
to handle this, each time an agent updates its value of $\clock$,
the remainder of its state is reset and it begins the rest of the protocol anew.
Restarting the downstream protocol is a known technique in population protocols also used in~\cite{GS18} to compose two leader elimination subprotocols.
The agents then generate $K$ additional geometric random variables in sequence, taking their sum. 
Upon completing the generation and propagation of the $K$'th number, 
the agent divides the sum by $K$ and stores the result in their output field.
Composition with a downstream protocol is as simple as letting that protocol be the last phase.
However, since our protocol has a positive probability of failure, 
this would translate to the downstream protocol as well.

The time is $O(\log^\timeExponent n)$ by the following rough analysis
(details follow).
We propagate $K$ numbers one after each other and for each epidemic $O(\log n)$ time is required. Since we set $k = O(\log n)$ then the protocol will take $O(\log^2 n)$ total time to complete.

\opt{sub}{\vspace{-0.2cm}}
\subsection{Formal specification of protocol} 
\label{subsec:protocol}


Our protocol uses uniform random bits in multiple places. 
We assume agents have access to independent uniformly random bits.
In the protocol, agents start by dividing in two groups of \s\ and \a. \a\ agents are responsible for the most part of the algorithm including generating geometric random variables and propagating their maximums while the \s\ agents only provide memory to store the sum of $K$ maximum geometric random variables. We split the state space such that \a\ agents and \s\ agents are responsible to store different variables.
The space multiplexing is a common approach used in population protocols to reduce the space complexity of the protocols~\cite{Alistarh:2018Rati}.

Agents initially have no role ($\x$),
and partition into roles via $\x,\x \to \a,\s$.
Since this takes $\Theta(n)$ time to complete,
we add transitions $\a,\x \to \a,\s$ and $\s,\x \to \s,\a$,
converging in $O(\log n)$ time, 
with the price of deviating from $\frac{n}{2}$ for each role. 
\opt{final}{It is proven in~\cite{doty2018efficientArxiv}}
\opt{full}{By Lemma~\ref{lem:boundOnCardinalityOfA}}
this deviation is $O(\sqrt{n \ln n})$,
increasing the size estimation error by merely a constant additive factor.\footnote{
    This mechanism of splitting the population approximately in two works for our protocol,
    because the number of $\a$ agents is likely to be so close to $n/2$ that our estimate of $\log n$ is reduced by an additive factor likely to be very close to $-1$.
}
All agents start at $\phase =0$.
The \a\ agents generate one geometric random variable 
(called \clock) and continue by propagating the maximum among the whole population. 
Since we use this \clock\ value for all early estimation of $\log n$, each time an agent finds out there was a greater value for the \clock\ than its own, it will reset all other computations that might have happened.

\opt{full}{By Lemma~\ref{lem:clkvalue}, the}
\opt{final,sub}{The}
maximum \clock\ amongst the population is a $2$ factor estimation of $\log n$
\opt{full}{\cite{doty2018efficientArxiv}}.
When any agent updates its $\clock$ with a new maximum,
it restarts the entire downstream protocol via $\reset$.
Once the maximum $\clock$ value is generated in the population,
it propagates (triggering $\reset$) by epidemic in $O(\log n)$ time.
The \clock\ variable could be used to estimate $K$, which is the number of independent additional geometric random variables each agent will generate. We also use \clock\ to set the leaderless phase clock inside each agent. In each epoch, the \a\ agents will generate one new geometric random variable and propagate its maximum. They count their number of interactions in each epoch using the \counter\ variable. At the end of an epoch, when \counter\ reaches $\cte.\clock$, the \a\ agents accumulate the value of the maximum \gr\ into the \ksum\ of a \s\ agent. The \a\ agents increase their \phase\ variable by one and set $\counter =0$ after either passing the geometric random variable to a \s\ agent or interacting with a \s\ agent in a higher phase. Separately, \s\ agents are responsible to propagate the maximum \ksum\ and maximum \phase\ among themselves.

In the \sizest\ protocol, all agents in role \a\ will finally generate $K =\constantNumGrv \cdot \clock$ geometric random variable and let the \s\ agents to store a sum of maximum one generated for each phase. 
Once all agents reach $\phase =\constantNumGrv \cdot \clock$ they set $\doneEST = \true$ and $\outpt = \frac{\ksum}{\phase}+1$.
We use $|\a|$, $|\s|$ for the cardinality of \a\ and \s\ agents respectively.
\begin{algorithm}[ht]
	\floatname{algorithm}{Protocol}
	\caption{$\sizest(\rec, \sen)$}
	\label{protocol:estimation}
	\begin{algorithmic}[100]		
        
		\LeftComment {initial state of agent:
		
		    $\role = \x,$
		    
    		$\counter = 0,		
    		\ksum = 0,		
    		\phase = 0,$
    		
    		$\gr = 1,
    		\clock = 1,$
		
		    
		    
		    $\doneEST = \false$
		}
		
    	\State{$\partitiontoAnS(\rec, \sen)$}
    	\If {$\rec.\role = \a $} 
    		\State{$\rec.\counter \gets \rec.\counter +1$}
		    \State{$\timerDone(\rec)$}
        \EndIf
        \If {$\sen.\role = \a $} 
    		\State{$\sen.\counter \gets \sen.\counter +1$}
    		\State{$\timerDone(\sen)$}
        \EndIf
        \State{$\epidemicClk(\rec, \sen)$}
        \State{$\epidemicPhase(\rec,\sen)$}
        \If{\text{one agent have $\role = \s$ and one have $\role = \a$}
        }
	        \State{$\updateSum(\rec, \sen)$}
        \EndIf
        \If{\text{both agents have $\role = \a$}}
            
		    \State{$\epidemicGr(\rec, \sen)$}
	    \EndIf
		\If{\sen.\doneEST}	
		    \State{$\outpt \gets \frac{\ksum}{\phase}+1
		        $}
	    \EndIf
	
	\end{algorithmic}
\end{algorithm}
\begin{algorithm}[!htbp]
	\floatname{algorithm}{Subprotocol}
	\caption{$\partitiontoAnS(\rec, \sen)$}
	\label{subprotocol:partition}
	\begin{algorithmic}[100]
	    \LeftComment{Partition the population in two almost equal size subpopulations.}
	    
        \If {$\sen.\role = \x , \rec.\role = \x$} 
            \State{$\sen.\role \gets \a$}
            \State{\text{$\sen.\clock \gets$ one geometric random variable}}
		    \State{$\rec.\role \gets \s$}
		\ElsIf{$\sen.\role = \a , \rec.\role = \x$}
		    \State{$\rec.\role \gets \s$}
		\ElsIf{$\sen.\role = \s , \rec.\role = \x$}
		    \State{$\rec.\role \gets \a$}
		    \State{\text{$\rec.\clock \gets$ one geometric random variable}}
		\EndIf
	\end{algorithmic}
\end{algorithm}


\begin{algorithm}[ht]
	\floatname{algorithm}{Subprotocol}
	\caption{$\epidemicClk(\agent1, \agent2)$}
	\label{subprotocol:epidemic_clk}
	\begin{algorithmic}[100]
	    \LeftComment{Maximum generated geometric variable for $\clock$ will be propagated.}
        \If {$\agent1.\clock < \agent2.\clock$} 
            \State{$\agent1.\clock \gets \agent2.\clock$}
		    \State{$\reset(\agent1)$}
		\ElsIf{$\agent2.\clock < \agent1.\clock$} 
            \State{$\agent2.\clock \gets \agent1.\clock$}
		    \State{$\reset(\agent2)$}
		\EndIf
	\end{algorithmic}
\end{algorithm}

\begin{algorithm}[!htbp]
 	\floatname{algorithm}{Subprotocol}
 	\caption{$\reset(\agent)$}
 	\label{subprotocol:reset}
 	\begin{algorithmic}[100]
 	\State {$\counter \gets 0$,		
	$\ksum \gets 0$,		
	$\phase \gets 0$}
    \State{$\gr \gets$ one geometric random variable}
	\State{$\doneEST \gets \false$}
 	    
 	     
 	      
		
 		
 	\end{algorithmic}
 \end{algorithm}
 
 
\begin{algorithm}[!htbp]
	\floatname{algorithm}{Subprotocol}
	\caption{$\epidemicGr(\agent1, \agent2)$}
	\label{subprotocol:epidemic_gr}
	\begin{algorithmic}[100]
	    \LeftComment{Maximum generated geometric variable for $\gr$ will be propagated.}
	    \If{$\agent1.\phase = \agent2.\phase$}
            \If {$\agent1.\gr < \agent2.\gr$} 
                \State{$\agent1.\gr \gets \agent2.\gr$}
    		\ElsIf{$\agent2.\gr < \agent1.\gr$} 
                \State{$\agent2.\gr \gets \agent1.\gr$}
    		\EndIf
		\EndIf
	\end{algorithmic}
\end{algorithm}

\begin{algorithm}[ht]
	\floatname{algorithm}{Subprotocol}
	\caption{$\timerDone(\agent)$}
	\label{subprotocol:timer_done}
	\begin{algorithmic}[100]
	    \LeftComment{Agents compare their \counter\ value to the specified threshold.}
        \If {$\agent.\counter = \cte\times \agent.\clock$, $\agent.\doneEST=\false$, $\agent.\passedGR=\true$} 
            \State{$\agent.\phase \gets \agent.\phase +1$}
            \State{$\movetonextGR(\agent)$}
        \EndIf
		\If{$\agent.\phase = \constantNumGrv \times \agent.\clock$} 
            \State{$\agent.\doneEST \gets \true$}
		\EndIf
	\end{algorithmic}
\end{algorithm}

\begin{algorithm}[ht]
	\floatname{algorithm}{Subprotocol}
	\caption{$\epidemicPhase(\agent1, \agent2)$}
	\label{subprotocol:epidemic_phase}
	\begin{algorithmic}[100]
	    \LeftComment{The maximum \phase\ will be propagated.}
	    \If{\text{both agents have $\role = \a$}}
            \If {$\agent1.\phase < \agent2.\phase$} 
                \State{$\agent1.\phase \gets \agent2.\phase$}
                \State{$\movetonextGR(\agent1)$}
            \ElsIf {$\agent2.\phase < \agent1.\phase$} 
                \State{$\agent2.\phase \gets \agent1.\phase$}
                \State{$\movetonextGR(\agent2)$}
            \EndIf
        \ElsIf{\text{both agents have $\role = \s$}}
            \If {$\agent1.\phase < \agent2.\phase$} 
                \State{$\agent1.\phase \gets \agent2.\phase$}
                \State{$\agent1.\ksum \gets \agent2.\ksum$}
            \ElsIf {$\agent2.\phase < \agent1.\phase$} 
                \State{$\agent2.\phase \gets \agent1.\phase$}
                \State{$\agent2.\ksum \gets \agent1.\ksum$}
            \EndIf
        \EndIf
	\end{algorithmic}
\end{algorithm}
 
\begin{algorithm}[!htbp]
	\floatname{algorithm}{Subprotocol}
	\caption{$\movetonextGR(\agent)$}
	\label{subprotocol:move_to_next_gr}
	\begin{algorithmic}[100]
            \LeftComment{The \agent\ move to the next \phase:}
		    
		    
            \State{$\agent.\counter \gets 0$}
            \State{$\agent.\gr \gets$ one geometric random variable}
            \State{$\agent.\passedGR = \false$}
            
	\end{algorithmic}
\end{algorithm}

\begin{algorithm}[!htbp]
	\floatname{algorithm}{Subprotocol}
	\caption{$\updateSum(\agent1, \agent2)$}
	\label{subprotocol:update_sum}
	\begin{algorithmic}[100]
            \LeftComment{The \agent\ accumulates the current value of \gr\ in \ksum:}
            \State{\text{a $ \gets$ agent with $\role =\a$}}
            \State{\text{s $ \gets$ agent with $\role =\s$}}
		    \If{a$.\phase =$ s$.\phase$,
		    a.$\counter \geq \cte \cdot $ a.$\clock$, 
		    and a.$\doneEST=\false$}
		            \State{s.$\phase \gets$ s.$\phase+1$}
		            \State{s.$\ksum \gets$ s.$\ksum + $ a.$\gr$}
		            \State{a.$\passedGR = \true$}
		    \ElsIf{a.$\phase < $ s.$\phase$}      
		        \State{a.$\passedGR = \true$}
		    \EndIf
	\end{algorithmic}
\end{algorithm}

\opt{sub}{The following corollary shows that close to half of agents end up in role $\a$. A generalized Lemma and its proof appears in the appendix.
}
\opt{full}{
    The following lemma shows that close to half of agents end up in role $\a$.
    
    \restateableLemma{lem:boundOnCardinalityOfA}{LemboundOnCardinalityOfA}{
        Let $a > 0$.
        In the \sizest\ protocol the cardinality of agents with \a\ role is in the interval of 
        $\left[ \frac{n}{2}-a, \frac{n}{2}+a \right]$
        with probability $\geq 1-e^{-2a^2/n}$.
    }{
        All agents in the \sizest\ protocol start in role \x. 
        In the \partitiontoAnS\ protocol agents will be assigned to their new roles. 
        Finally, all agents participate in the \sizest\ protocol either having role \a\ or \s. 
        Thus, after completion of the \partitiontoAnS\ protocol $|\a|+|\s| = n$ holds. 
        
        The percentage of agents that change to role $\a$ is an average of the percentage 
        of $\a$'s produced by the first rule (always exactly $1/2$)
        and the percentage produced by the next two rules.
        The next two rules ensure that if the percentage of $\a$'s so far produced is greater than $1/2$,
        then $\a$ is less likely to be produced next than a fair coin flip
        (since, conditioned on the next interaction being between one $\x$ and one non-$\x$,
        the probability of producing $\a$ is exactly $\frac{|\s|}{|\a|+|\s|}$,
        i.e., smaller if there are more $\a$'s than $\s$'s),
        and vice versa.
        Thus, 
        the distribution of the percentage difference 
        $\left| \frac{|\a|}{n} - 1/2 \right|$ 
        of $\a$'s is stochastically dominated by the difference between the percentage of heads of a fair-coin binomial distribution $\mathbf{B}(n,1/2)$ and $\mathbf{B}$'s expected percentage of $1/2$.
        Therefore we can use a binomial distribution to bound the upper and lower tails of the distribution of the number of eventual $\a$'s.
        For any $a$ the Chernoff bound says for any $a > 0,$
        $\Pr{\mathbf{B}(n, 1/2) \geq n/2 + a} \leq e^{-2a^2/n}$
        and 
        $\Pr{\mathbf{B}(n, 1/2) \leq n/2 - a} \leq e^{-2a^2/n}$.
    }
    
    \proofLemboundOnCardinalityOfA
}

\begin{corollary}\label{cor:lazyboundOnCardinalityOfA}
    In the \sizest\ protocol the cardinality of agents with \a\ role is in the interval of 
    $\left[ \frac{n}{3}, \frac{2n}{3} \right]$
    with probability $\geq 1-e^{-n/18}$.
\end{corollary}

In each epoch, one geometric random variable 
(in the first epoch \clock\ and in the subsequent epochs \gr) 
is generated and its maximum will be propagated by epidemic among the population. We set the time of each epoch equal to the required time of generating one plus the time for completion of an epidemic.
To analyze the time complexity of our protocol, 
we require the time bounds for completing an epidemic from the paper~\cite{AAE08}. 
The current form is taken from~\cite{doty2018exact}.
For all $n \in \N^+$,
let $H_n = \sum_{k=1}^n \frac{1}{k}$ denote the $n$'th harmonic number.
Note that $\ln n \leq \frac{n-1}{n} H_{n-1} \leq 1 + \ln n$.

\opt{final}{
\begin{lemma}[\cite{AAE08}]\label{lem:epidemic}
Let $T$ denote the time to complete an epidemic.
Then
$\E{T} = \frac{n-1}{n} H_{n-1}$,
$\Pr{T < \frac{1}{4} \ln n} < 2 e^{-\sqrt{n}}$,
and for any $\alpha_u > 0$,
$\Pr{T > \alpha_u \ln n} < 4 n^{- \alpha_u/4+1}$.
\end{lemma}
}

The following corollary describes an epidemic in a subpopulation.
This refers to some subset $S$ of the population executing epidemic transitions only among themselves,
which slows down the epidemic by only a constant factor if $|S| = \Omega(n)$.

\opt{full}{
\restateableCorollary{cor:partialEpidemic}{CorPartialEpidemic}{
    Let $c \geq 1$.
    Suppose an epidemic happens among a subpopulation of $a = n/c$ agents.
    Let $\mathbf{T}$ denote the time to complete such an epidemic.
    Then
    for any $\alpha_u > 0$,
    $\Pr{\mathbf{T} > \alpha_u \ln a} < a^{ -(\alpha_u-4c)^2 / 12c }$.
    }
    {
    The probability that in the next interaction the scheduler picks two agents from the subpopulation $a$ is $\binom{a}{2} / \binom{n}{2} = \frac{a(a-1)}{n(n-1)}$. 
    By Lemma~\ref{lem:epidemic}, 
    if $\mathbf{T}'$ denotes the time to complete an epidemic in population size $a$, 
    then $\E{\mathbf{\mathbf{T}}'} = \frac{a-1}{a} H_{a-1}$. 
    Since we have $\frac{n(n-1)}{a(a-1)}$ expected interactions in the whole population of size $n$
    in order to obtain one interaction within the subpopulation,
    the expected time to complete this epidemic (counting total interactions in the whole population)
    is $\E{\mathbf{T}} = \frac{n(n-1)}{a(a-1)} \cdot \E{\mathbf{T}'} = \frac{n(n-1)}{a(a-1)} \frac{a-1}{a}H_{a-1} = \frac{n(n-1)}{a^2}H_{a-1} \geq c^2 \ln a.$ 
    By the Chernoff bound we have:
    \begin{eqnarray*}
        \Pr{\mathbf{T} \geq (1+\delta)\E{\mathbf{T}}}
    &\leq&
        e^{-\delta^2 \E{\mathbf{T}} / 3}
    \\&\leq&
        e^{-\delta^2 \cdot c^2 \ln(a) / 3}
    \\&=&
        a^{-c^2\delta^2 / 3} 
    \end{eqnarray*}
    Setting $\alpha_u = 4c(1+\delta)$, 
    $\Pr{\mathbf{T} \geq \alpha_u \ln a} \leq
    a^{ -((\alpha_u/4c)-1)^2c^2 / 3 }$.
}
\proofCorPartialEpidemic
}
Setting $c=3$ and $\alpha_u = 24$ in 
\opt{final}{Corollary A.3 of~\cite{doty2018efficientArxiv}}
\opt{full}{\ref{cor:partialEpidemic}}
gives the following.
\begin{corollary}\label{cor:partialEpidemic-c-equals-3}
    Suppose an epidemic happens among a subpopulation of $n/3$ agents with time $\mathbf{T}$.
    Then $\Pr{\mathbf{T} > 24 \ln n} < 27 n^{-3}$.
\end{corollary}



The next lemma bounds the number of interactions an agent has in a given time,
and it is the basis of the leaderless phase clock we use.
\opt{sub}{It is proven in the appendix.}
\opt{final}{It is proven in~\cite{doty2018efficientArxiv}.}
It follows from a simple Chernoff bound on the number of interactions involving a single agent in a given window of time.

\restateableLemma{lem:boundONepoch}{LemBoundOnepoch}{
    Let $C \geq 3$ and $D = 2C + \sqrt{12 C}$.
    In time $C \ln n $, with probability $\geq 1-1/n$, 
    each agent has at most $D \ln n$ interactions.
}{
    Fix an agent $a$.
    Let $\mathbf{I}_a$ be the number of interactions involving $a$ during  $C n\ln n$ total interactions ($C \ln n$ time).
    The probability that any given interaction involves $a$ (either receiver or sender) is exactly $\frac{2}{n}$, 
    so $\mathbf{I}_a$ is distributed binomially, and
    $\E{\mathbf{I}_a} = C n \ln n \cdot \frac{2}{n} = 2 C \ln n$. 
    Applying the Chernoff bound,
    for any $0 < \delta \leq 1$,
    \[
        \Pr{\mathbf{I}_a \geq (1+\delta) 2C\ln n} 
    \leq
        e^{-\delta^2 2C\ln(n) / 3}
    =
        n^{-2C\delta^2 / 3}.
    \]
    Let $D = 2C + \sqrt{12 C}$,
    and let $\delta = \frac{D}{2C} - 1$.
    (Note $\delta \leq 1$ so long as $C \geq 3$.)
    Then $(1+ \delta) 2C \ln n = D \ln n$
    and
    $n^{-2C\delta^2 / 3} = n^{\frac{-(D-2C)^2}{6C}} = n^{-2}$.
    Then 
    $
        \Pr{\mathbf{I}_a \geq D \ln n} \leq
    n^{-2}.
    $
    By the union bound,
    $
        \Pr{(\exists a)\ \mathbf{I}_a \geq D\ln n} 
    \leq
        n^{-1}.
    $
    So, by setting $D = 2C+ \sqrt{12C}$ we bound the probability that each agent has more than $D \ln n$ interactions in time $C \ln n $ to be $\leq 1/n$.
}

\opt{full}{\proofLemBoundOnepoch}

\begin{corollary}\label{cor:exactboundonepoch}
Each agent has $\geq 65 \ln n$ interactions
in time $24 \ln n$
with probability $\leq 1/n$.
\end{corollary}

By Lemma~\ref{lem:boundONepoch} each agent has at most 
$\left(2\cdot 24 +\sqrt{12 \cdot 24}\right) \ln n \leq 65 \ln n \leq 94 \log n$ interactions in the time that it takes to generate and propagate maximum of one geometric random variable. 
Thus, each agent should count up to $94 \log n$ for its leaderless phase clock,
to ensure that with high probability none reaches that count until the maximum geometric random variable is known to all agents. 
However, agents are not aware of any prior approximation of $\log n$. In the \sizest\ protocol, agents use their \clock\ variable for this approximation. 
As mentioned, all the agents in role \a\ start by generating one geometric random variable \clock.
The maximum in the population is used as a weak (constant factor) approximation of $\log n$. 
\opt{full}{
Corollary~\ref{cor:half-geom-tighter-bounds} says that the maximum of $|\a|$ geometric random variables is in the interval of
$[\log |\a| - \log \ln |\a|, 2\log |\a|]$ 
with probability at least $1-1/|\a|$. However, we are using the \clock\ and \gr\ variables as an approximation of $\log n$ rather than $\log |\a|$. Lemma~\ref{lem:clkvalue} will give us a bound over the \clock\ value with respect to $\log n$. Corollaries~\ref{cor:grvalue},~\ref{cor:boundONtime},~\ref{cor:boundONcoefficient} use this lemma for a bound over \gr, \counter, and \phase\ values.
}
\opt{sub}{Their statement and proofs appear in the appendix.}
\opt{final}{Their statement and proofs appear in~\cite{doty2018efficientArxiv}.}
\opt{full}{
\begin{lemma}\label{lem:clkvalue}
    The \clock\ value generated by \genClk\ is in the interval of $[\log n - \log \ln n, 2\log n+1]$ with probability at least $1-1/n-e^{-n/18}$.
\end{lemma}
\begin{proof}
    
    By Corollary~\ref{cor:lazyboundOnCardinalityOfA}, $n/3\leq |\a| \leq 2n/3$ with probability $\geq 1-e^{-n/18}$. We apply the result of Corollary~\ref{cor:half-geom-tighter-bounds} and substitute $n$ with $|\a|$:
    \begin{alignat*}{2}
    \log {(n/3)} - \log \ln {(n/3)}
    &\leq 
    \clock 
    &\leq 
    2\log {(2n/3)}\\
    \log n -1.6 - \log \ln n -0.4
    &\leq 
    \clock 
    &\leq 
    2\log n + 2\log (2/3)\\
    \log n - \log \ln n -2
    &\leq 
    \clock 
    &\leq 
    2\log n-1\\
    \log n - \log \ln n -2
    &\leq 
    \clock 
    &\leq 
    2\log n-1
    \end{alignat*}
    So, by setting $\clock = \clock +2$ for all agents, this variable is in the interval of $[\log n -\log \ln n , 2\log n + 1]$ with probability $\geq 1-1/n-e^{-n/18}$.
\end{proof}
}


\opt{final}{
    \begin{corollary}\label{cor:grvalue}
    The \clock\ (\gr) value generated by \genClk\ (\genGr) is in the interval of $[\log n - \log \ln n-2, 2\log n-1]$ with probability at least $1-1/n-e^{-n/18}$.
    \end{corollary}
    
    \begin{corollary}\label{cor:boundONtime}
    The number of interactions in each epoch in the $\sizest$ is in the interval 
    $[\cte \log n - \cte \log \ln n, 189\log n]$ with probability $\geq 1-1/n-e^{-n/18}$. \end{corollary}
    
    \begin{proof}
    By Corollary~\ref{cor:exactboundonepoch}, agents should count up to $96\ln \leq 139 \log n$ before moving to the next epoch. if we set the threshold of the \counter\ to $\cte \cdot \clock$, $\cte \log n - \cte  \log \ln n \geq 93 \log n$ then the \counter\ variable will be in the interval of $[\cte \log n - \cte \log \ln n, 188\log n+\cte]$ with high probability ($188 \log n + \cte \leq 189 \log n$ for $n \geq 2$).
    \end{proof}
    
    \begin{corollary}\label{cor:boundONcoefficient}
    The number of epochs in the $\sizest$ is in the interval 
    $[\constantNumGrv\log n - \constantNumGrv\log \ln n, 11\log n]$ with probability $\geq 1-1/n-e^{-n/18}$. \end{corollary}
    
    \begin{proof}
    By Corollary C.10 in~\cite{doty2018efficientArxiv},
    to achieve the additive error of \logSizeError\ for our protocol the number of geometric random variables should be $\geq 4\log n$. By setting the threshold of the number of phases to $\constantNumGrv \times \clock$, for $n\geq 200$, $\constantNumGrv \log n - \constantNumGrv \log \ln n \geq 4\log n$. The number of phases will be in the interval of $[\constantNumGrv\log n - \constantNumGrv\log \ln n, 10\log n + \constantNumGrv]$ with high probability ($10 \log n + \constantNumGrv\leq 11 \log n$ for $n\geq 2$).
    \end{proof}
}

The next Lemma bounds the space complexity of our main protocol
by counting the likely range taken by the variables in $\sizest$.
\opt{sub,final}{
    It is proven 
    \opt{sub}{in the Appendix.}
    \opt{final}{in~\cite{doty2018efficientArxiv}.}
}

\restateableLemma{lem:space}{LemSpace}{
    $\sizest$ uses $O(\log^4 n)$ states with probability $\geq 1-O(\log n)/n$.
}{
    With probability at least $1 - O(1/n)$
    (see individual lemma statements for constants in the $O$),
    the set of values possibly taken on by each field are given as follows:
    \begin{center}
    \begin{tabular}{ l l l }
     \clock & $\{1, \ldots, 2\log n+1\}$ 
            & Lemma~\ref{lem:clkvalue} \\ 
     \gr & $\{1, \ldots, 2\log n\}$ 
            & Corollary~\ref{cor:grvalue} \\ 
     \counter & $\{0, \ldots, 191 \log n\}$ 
            & Corollary~\ref{cor:boundONtime} \\
     \phase & $\{0,\ldots, 11 \log n\}$ 
            & Corollary~\ref{cor:boundONcoefficient} \\
     \ksum & $\{0, \ldots, 22 \log ^2 n\}$ 
            & Corollaries~\ref{cor:grvalue},~\ref{cor:boundONcoefficient}
    \end{tabular}
    \end{center}
    
    In our protocol we used space multiplexing to reduce the number of states agents use. 
    The \a\ agents are responsible to generate geometric random variables and propagate the maximum among themselves. Thus, they store \clock, \gr, \counter, and \phase\ variables. 
    While the \s\ agents are only responsible to hold the sum of all geometric maximas and they store \clock, \phase, and \ksum. After each agent sets $\doneEST =\true$, 
    it no longer needs to store the value in \gr\ or \phase\ and can use that space to store the result of $\ksum / \phase +1$ as the output. Although we are using the explained space multiplexing to reduce the number of states used by the agents, both \a\ and \s\ agents need to store \clock\ and \phase\ to stay synchronized. 
    Note that the probability that each geometric random variable is greater than $2\log n$ is less than $1/n$, by the union bound the probability that any of them is greater than $2\log n$ is less than $\frac{11\log n}{n}$.
}

\opt{full}{\proofLemSpace}

The next corollary bounds the time complexity of protocol \sizest;
the main component of the time complexity is that $\Theta(\log n)$ geometric random variables must be generated and propagated by epidemic among the population,
each epidemic taking $\Theta(\log n)$ time.
\opt{sub}{A proof appears in the appendix.}
\opt{final}{A proof appears in~\cite{doty2018efficientArxiv}.}
\restateableCorollary{cor:timecomplexity}{CorTimeComplexity}{
    The \sizest\ protocol converges in $O(\log^2 n)$ time with probability at least $1-1/n^2$.
}{
    By Corollary~\ref{cor:partialEpidemic-c-equals-3}, with probability $\geq 1-(27/n^3)$ propagating the maximum of the \clock\ variable takes at most $24\ln n$ time.
    
    By Corollary~\ref{cor:boundONcoefficient}, at most $11 \log n$ geometric random variables will be generated, 
    and by Corollary~\ref{cor:partialEpidemic-c-equals-3}, with probability $\geq 1-(27/n^3)$ a given variable takes at most $24\ln n$ time to propagate its maximum (total of $11 \log n\cdot 24 \ln n$ time). Note that, it takes constant time for \a\ agents to interact with a \s\ agent and move to the next \phase.
    
    By the union bound over all epochs, 
    the probability that generating $11 \log n +1 $ geometric random variables and propagating their maximum takes more than $(11\log n +1) \cdot 24 \ln n$ time is $\geq 1- \frac{O(\log n)}{n^3} \geq 1-1/n^2$ for large values of $n$.
}
\opt{full}{\proofCorTimeComplexity}

The following result is a Chernoff bound on sums of random variables, each of which is the maximum of independent geometric random variables (with probability of success $\frac{1}{2}$).
\opt{full}{It is a corollary of Corollary~\ref{cor:sumofgeomConcreteK}, proven in the appendix.}
\opt{final}{It is a corollary of a similar Chernoff bound proven in the appendix of~\cite{doty2018efficientArxiv}.}

\restateableLemma{lem:sumofgeometricamongAagents}{LemSumOfGeometricAmongAagents}{
    Let $K \geq 4\log n$ and $a$ be a number in the interval of $[n/2 -\sqrt{n \ln n}, n/2+\sqrt{n\ln n}]$. 
    Let $\ksum/K$ be the average of $K$ $\frac{1}{2}$-geometric random variables.
    Then $\Pr{ \left| \frac{\ksum}{K}+1 - \log n \right| \geq \logSizeErrorAagents } 
        \leq \frac{6}{n}$.
}{
        By Corollary~\ref{cor:sumofgeomConcreteK}, 
    $
            \Pr{ \left| \frac{\ksum}{K} - \log a \right| \geq \logSizeError } 
        \leq 
            \frac{2}{a}.
    $
     Since $n/2 -\sqrt{n \ln n} \leq a \leq  n/2+\sqrt{n\ln n}$ then $\log n -2 \leq \log a \leq \log n$. Hence:
     $
            \Pr{ \left| \frac{\ksum}{K}+1 - \log n \right| \geq \logSizeError+1 } 
        \leq 
            \frac{2}{a} \leq \frac{6}{n}.
     $
    }

\opt{full}{
    \proofLemSumOfGeometricAmongAagents
}

\restateableLemma{lem:correctnessofProtocol}{Lemcorrectness}{
    In the \sizest\ protocol, with probability $1$, all agents converge to the same value $C$ in their \outpt\ field.
    Furthermore, 
    $\Pr{ \left| C - \log n \right| \geq \logSizeErrorAagents } 
        \leq 9/n$.
}{  
    The convergence of all agents to a common value of $C$ with probability 1 is evident from inspection of the protocol.
    The \sizest\ protocol calculate $\log n$ within an additive error of $5.7$ if:
    \begin{itemize}
        \item 
        The \clock\ variable generated at the beginning of the protocol is $\geq \log n -\log \ln n$ (not too small). By Lemma~\ref{lem:clkvalue}, the value of \clock\ can be less than $\log n - \log  \ln n$ with probability at most $1/n + e^{-n/18}$.
        
        \item 
        The number of agents in role \a\ is less than $n/2-\sqrt{n\ln n}$ or greater than $n/2+\sqrt{n\ln n}$. By Lemma~\ref{lem:boundOnCardinalityOfA}, $    a\leq  \frac{n}{2}-\sqrt{n\ln n}$ or $ a \geq  \frac{n}{2}+\sqrt{n\ln n}$ with probability at most $e^{-2\ln n} = 1/n^2$.
        
        \item 
        The \a\ agents propagate each geometric random variables among themselves within $24\ln n$ time. By Corollary~\ref{cor:partialEpidemic-c-equals-3}, an epidemic might take more than $24\ln n$ time with probability at most $\frac{27}{n^3}$. In the \sizest\ protocol there are $O(\log n)$ epidemics in total. Thus, by the union bound the probability that any of them take more than $24 \ln n$ time is at most $O(\log n)/n^3 \leq 1/n^2$ for large values of $n$.
        
        \item 
        An \phase\ terminates before completion of one epidemic. This can happen if one agent has too many interaction in an \phase. By Corollary~\ref{cor:exactboundonepoch}, for all agents, the probability that any of them have more than $65 \ln n$ interaction in $24 \ln n$ time is $\leq 1/n$. 
        
        \item 
        By Lemma~\ref{lem:sumofgeometricamongAagents}, the average of $K$ geometric random variables among \a\ agents might be out of the interval of $\left[\log n - 5.7, \log n + 5.7\right]$ with probability at most $6/n$.
    \end{itemize}
    By the union bound, the probability that \outpt\ reports a value with an additive error more than $5.7$ is less than $8/n + 2/n^2 + e^{-n/18} \leq 9/n$
}

\opt{full}{
    \proofLemcorrectness
}
    
\opt{full}{
    Finally, we combine these results to prove the main result of this section, Theorem~\ref{thm:main-size-estimation-protocol}.
    \proofTHMmainSizeEstimationProtocol
}

\opt{sub,final}{
    
    \noindent
    {\bf Termination with a leader and guaranteed size upper bound.}
    Two other results are discussed in more detail
    \opt{sub}{in the appendix.}
    \opt{final}{in~\cite{doty2018efficientArxiv}.}
    The first is that we can make the size-estimation protocol terminating with high probability using an initial leader.
    Intuitively,
    the leader can be used to trigger an epidemic-based phase clock used to count to $\Omega(\log^2 n)$,
    enough time for the protocol to probably have converged.
    The second discusses the possibility of transforming the size estimation,
    which has a small probability of being much larger \emph{or} much smaller than the actual size,
    into a \emph{guaranteed upper bound} on the population size.
}

\noindent
    {\bf Reducing the space complexity.}
    In our protocol, we used space multiplexing, a known technique in population protocols that split the state space such that different agents are responsible to store different variables. 
    Although this technique reduces the number of states per agent, 
    we cannot push it further with the current scheme. 
    Our protocol is dependent on \emph{all} agents agreeing on the values of \clock\ and \phase\ to stay synchronized. 
    Thus, if an agent participates in the protocol it is required to stores the updated value of both \clock\ and \phase.

\opt{full}{
    \subsection{Probability-1 estimation of upper bound on $\log n$}
It is not clear how to make our main protocol correct with probability 1,
meaning that it guarantees the estimate $k$ obeys $|k - \log n| \leq \logSizeErrorAagents$.
The protocol could err in either direction and make $k$ too large or too small, 
depending on the sampled values of the geometric random variables.

However, for many applications using an estimate of $\log n$, 
an \emph{upper} bound is sufficient to ensure correctness
(though being too large may slow things down).
A straightforward modification of our protocol guarantees that $k \geq \log n$ with probability $1$,
while preserving the high-probability asymptotic time complexity.
We run a slow, exact backup protocol that stabilizes to $\ke$ such that $2^{\ke-1} < n \leq 2^{\ke}$.
This is accomplished by transitions $\ell_i,\ell_i \to \ell_{i+1},f_{i+1}$ for all $i$ 
and $f_i,f_j \to f_i,f_i$ for all $j < i$,
where all agents start with $\ell_0$.
After $O(n)$ time all agents store $\ke$ in their subscript.
Note this approaches $\ke$ from below.
Then modify our main protocol estimating $\log n$ to add $3.7$ (with probability $ \geq 1-2 \cdot  e^{\kappa /2 - t/4}$ where $\kappa$ is the number of geometric random variables; by Lemma~\ref{lem:known_bound_for_expected_value_of_max}, and setting $\alpha  =1$ ) in Lemma~\ref{lem:chernoff-geometric} since we can leverage the corollary~\ref{cor:half-geom-is-sub-exp} for one side error)
to its estimate of $\log n$, calling the result $k$;
with high probability $k \geq \log n$.
We then get a guaranteed upper bound on $\log n$ by reporting $\max(k,\ke)$ at any moment:
the former converges to $k \geq \log n$ with probability of failure $O(\log(n) / n)$.
If this fails (i.e., if $k < \log n$),
the value $\ke$ is guaranteed eventually to exceed $k$.
The contribution to the expected time of the latter case is negligible,
so the expected convergence time remains $O(\log^2 n)$.

Note that $k$ may exceed $\log n$ by an arbitrary amount,
with low probability.
In the terminology of Section~\ref{sec:correct-defn-variations},
we have changed the definition of ``correct'' from 
``$|k = \log n| \leq \logSizeErrorAagents$'' 
to ``$k \geq \log n$'',
and showed probability 1 of correctness under the new definition.
(Note that we still guarantee that $k \leq \log n + O(1)$ with high probability, where the $O(1)$ constant is 
now $\logSizeErrorAagents + 3.7 = 9.4$.)
An interesting open question is to find a polylogarithmic protocol that guarantees $k$ is within $O(1)$ of $\log n$ with probability 1.

    \subsection{Terminating size estimation with a leader}
It is possible to make the size-estimation protocol terminating if we start with an initial leader.
By Theorem~\ref{thm:main-termination}, 
a leader
(or a $o(n)$-size junta of leaders) 
is \emph{required} for termination to work with positive probability.

\begin{theorem} \label{thm:size-estimation-terminating-with-leader}
    There is a uniform terminating population protocol with an initial leader
    that, with probability 
    $\geq 1 - O(\log n)/n$,
    computes and stores in each agent an integer $k$ such that 
    $|k - \log n| \leq \logSizeErrorAagents$,
    taking  
    $O(\log^\timeExponent n)$
    time
    and
    $O(\log^\numStatesExponent n)$
    states.
\end{theorem}

\begin{proof}
In the presence of a leader, the population can simulate a phase clock as described in Angluin et al.~\cite{AAE08}. 
Let the leader role be \a. By~\cite[Corollary 1]{AAE08}, there is a constant $k_1 = \max(8c,\frac{8d}{c})$ for the number of phases that it takes at least $d \ln n$ to reach phase $k_1$ with probability at least $1-\frac{1}{n^c}$. If we set the number of phases in a phase clock greater than $288$, then reaching the maximum phase takes at least $36 \ln n$ time with probability at least $1-\frac{1}{n}$. By Lemma~\ref{lem:epidemic}, $36 \ln n$ time is sufficient to generate and propagate the \clock\ variable. 
By setting the number of phases equal to $k_2 \cdot \constantNumGrv\cdot\clock$, we can set a timer to count up until $\frac{k_2}{8}\ln n \log n$ time with probability at least $1-\frac{1}{n}$ for some ``big'' $k_2$~\cite[Corollary 1]{AAE08}. When the phase clock reaches $k_2 \cdot \constantNumGrv\cdot\clock$, leader terminates stage and report the \outpt\ value it computed.
\end{proof}

}

\opt{sub}{\vspace{-0.2cm}}
\section{Termination} \label{sec:termination}

The concept of termination has been referenced and studied in population protocols~\cite{AR07, MCS12c, M15}, 
but to our knowledge no formal definition exists.
We give an abstract definition capturing the behavior of most protocols that ``perform a computational task''. 

Let $P$ be a protocol with a set $I$ of ``valid'' initial configurations, 
where each agent's memory has a Boolean field \terminated\ set to \false\ in every configuration in $I$.\opt{full}{\footnote{In the language of states, 
    we partition the state set $\Lambda$ into disjoint subsets 
    $\Lambda_T$ and $\Lambda_N$ such that 
    $\Lambda_T \cup \Lambda_N = \Lambda$
    and
    $\Lambda_T$ are precisely the states with $\terminated = \true$.
}}
A configuration $\vec{c}$ of $P$ is \emph{terminated} if 
at least one agent in $\vec{c}$ has $\terminated=\true$.
(Note the distinction with a \emph{silent} configuration, where no transition can change any agent's state~\cite{beauquier1999memory}.)
Let 
$\kappa > 0$ and 
$t:\N\to\N$.
$P$ is \emph{$\kappa$-$t$-terminating} if, 
for all $\vi \in I$, 
with probability $\geq \kappa$,
$P$ reaches from $\vi$ to a terminated configuration $\vec{c}$,
but takes time $\geq t(n)$ to do so.

This definition leaves totally abstract which particular task
(e.g., leader election)
is assumed to have terminated.
The idea is that if the task will not be complete before time $t(n)$ with high probability,
then no agent should set $\terminated=\true$ until time $\geq t(n)$ with high probability.
So proving an upper bound on $t(n)$ in the definition of terminating 
implies that no protocol can be terminating if it requires time $> t(n)$ to converge.

The definition is applicable beyond the narrow goal of terminating a population protocol.
It says more generally that a ``signal'' is produced after some amount of time.
This signal may be used to terminate a protocol,
move it from one ``stage'' to another,
or it may be some specific Boolean value relevant to a specific protocol,
where in any case the value will start $\false$ for all agents 
and eventually be set to $\true$ for at least one agent.


Let $\alpha>0$.
We say a configuration $\vc$ is \emph{$\alpha$-dense} if,
for all $s \in \Lambda$, $\vc(s) > 0 \implies \vc(s) \geq \alpha n$.
(Recall $n = \| \vc \|$.)
In other words, every state present occupies at least fraction $\alpha$ of the population.
We say protocol $P$ with valid initial configuration set $I$ is \emph{i.o.-dense} if there exists $\alpha>0$ such that infinitely many $\vi \in I$ are $\alpha$-dense.
In particular,
an i.o.-dense protocol does not always have an initial \emph{leader}:
a state present in count 1 in every $\vi \in I$.


The next theorem, our second main result, 
shows that termination is impossible for uniform i.o.-dense protocols that require more than constant time,
no matter the space allowed.

\begin{theorem}\label{thm:main-termination}
    Let 
    $\kappa > 0$ 
    and
    $t:\N\to\N$,
    and let $P$ be a uniform i.o.-dense population protocol.
    If $P$ is $\kappa$-$t$-terminating,
    then $t(n) = O(1)$.
\end{theorem}

Let $\Lambda$ be the  (possibly infinite) set of all states of a population protocol.
Recall the definition of randomized transitions from Section~\ref{sec:model};
We now introduce extra notation that will be useful in this Section.
We consider a transition \emph{relation} $\Delta \subseteq \Lambda^4$,
writing $a,b \to c,d$ to denote that $(a,b,c,d) \in \Delta$
(i.e., if agents in states $a$ and $b$ interact,
then one of the possible random outcomes is to change to states $c$ and $d$).
For $\rho \in (0,1]$,
we write $a,b \rxn^\rho c,d$ to denote that when states $a$ and $b$ interact,
with probability $\rho$ they transition to $c$ and $d$.
Say that $\rho$ is the \emph{rate constant} of transition $a,b \rxn^\rho c,d$.
If there exist $a,b \in \Lambda$ and $\rho' \geq \rho$ such that $a,b \rxn^{\rho'} c,d$,
we write $c \in \mathrm{PROD}_\rho(a,b)$ and $d \in \mathrm{PROD}_\rho(a,b)$.
(In other words, $c \in \mathrm{PROD}_\rho(a,b)$ if $c$ is produced with probability at least $\rho$ whenever $a$ and $b$ interact).
For any $\Gamma \subseteq \Lambda$ and $\rho \in [0,1]$, define
$
    \mathrm{PROD}_\rho(\Gamma) =  \{s \in \Lambda \mid (\exists a,b \in \Gamma)\ s \in \mathrm{PROD}_\rho(a,b) \}.
$\opt{full}{\footnote{
    In other words,
    $\mathrm{PROD}_\rho(\Gamma)$ is the set of states producible by a single transition, 
    assuming that only states in $\Gamma$ are present,
    and that the only transitions used are those that have probability at least $\rho$ of occurring when their input states interact.
}}

Let $\Lambda^0 \subseteq \Lambda$.
For $i \in \N^+$, define
$
    \Lambda^i_\rho = \Lambda^{i-1}_\rho \cup \mathrm{PROD}_\rho(\Lambda^{i-1}_\rho).
$
For $m \in \N$,
if $s \in \Lambda^m_\rho$,
we say $s$ is \emph{$m$-$\rho$-producible from $\Lambda^0$}.\opt{full}{\footnote{
    If $s$ is $m$-$\rho$-producible from $\Lambda^0$,
    then in other words,
    $s$ is producible from any sufficiently large configuration that contains only states in $\Lambda^0$,
    using at most $m$ different \emph{types} of transitions,
    each of which has probability at least $\rho$.
    More than one instance of each transition,
    however, 
    may be necessary.
    For instance, 
    with transitions $x_i,x_i \rxn^\rho x_{i+1},q$ for all $i \in \N^+$,
    $x_m$ is $m$-$\rho$-producible from $\Lambda_0=\{x_1\}$,
    but $2^m$ transitions of type $x_1,x_1 \to x_2,q$ must be executed,
    followed by $2^{m-1}$ of type $x_2,x_2 \to x_3,q$, etc.
}}
For configuration $\vc$,
we say $s$ is \emph{$m$-$\rho$-producible from $\vc$}
if $s$ is $m$-$\rho$-producible from $\Lambda^0 = \{ s \in \Lambda \mid \vc(s) > 0 \}$,
the states present in $\vc$.\opt{full}{\footnote{
    Note that $s$ may be $m$-$\rho$-producible from $\vc$,
    but not actually producible from $\vc$,
    if the counts in $\vc$ are too small for the requisite transitions to produce $s$.
}}

Our main technical tool is the following lemma,
a variant of the ``timer/density lemma'' of~\cite{Do14}
(see also~\cite{alistarh2017time}).
The original lemma states that in a protocol with $O(1)$ states,
from any sufficiently large $\alpha$-dense configuration,
in $O(1)$ time all states appear with $\delta$-density 
(for some $0 < \delta < \alpha$).
The proof is similar to that of~\cite{Do14},
but is re-tooled to apply to protocols with a non-constant set of states
(also to use the discrete-time model of population protocols,
instead of the continuous-time model of chemical reaction networks).\footnote{
    Alistarh et al.~\cite{alistarh2017time} also prove a variant applying to protocols with $\omega(1)$ states,
    but for a different purpose:
    to show that all states in $\Lambda$ appear as long as $|\Lambda| \leq \frac{1}{2} \log \log n$.
    However,
    beyond that bound,
    the lemma does not hold~\cite{GS18}.
    In our case,
    we are not trying to show that all states in $\Lambda$ appear,
    only those in some constant size subset of states,
    all of which are $m$-$\rho$-producible from the initial configuration.
}
The key new idea is that,
even if a protocol has infinitely many states
(of which only finitely many can be produced in finite time),
for any subset of states $\Lambda^m_\rho$ 
``producible via only $m$ transitions, 
each having rate constant at least $\rho$'',
all states in $\Lambda^m_\rho$ are produced in constant time with high probability from sufficiently large configurations.

\newcommand{\timeRand}[2]{\mathbf{C}_{#1,#2}}

\restateableLemma{lem:timer}{LemTimer}{
    Let $\alpha > 0$, $m \in \N^+$, $\rho \in (0,1]$,
    and $P$ be a population protocol.
    Then there are constants $\epsilon,\delta,n_0 > 0$ such that, 
    for all $n \geq n_0$,
    for all $\alpha$-dense configurations $\vc$ of $P$ with $n = ||\vc||$,
    the following holds.
    Let $\Lambda^m_\rho$
    be the set of states $m$-$\rho$-producible from $\vc$. 
    For $s\in \Lambda$ and $t>0$,
    let $\timeRand{t}{s}$ be the random variable denoting the count of $s$ at time $t$,
    assuming at time $0$ the configuration is $\vc$.
    Then
    $
        \Pr{(\forall s \in \Lambda^m_\rho)\ \timeRand{1}{s} \geq \delta n} 
        \geq 1 - 2^{-\epsilon n}.
    $
}{
    We need that $|\Lambda^m_\rho| < \infty$.
    To see this holds,
    assume otherwise.
    Note that if all pairs of states $a,b$ have a finite number of transitions of the form $a,b \to \ldots$,
    then this implies by induction that each $\Lambda^m_\rho$ is finite.
    So assume there are $a,b\in\Lambda$ and an infinite set of transitions $a,b \rxn^{\rho_i} \ldots$ for $i \in \N$.
    Because these are probabilities, 
    $\sum_{i=0}^\infty \rho_i \leq 1$.
    Then $\lim\limits_{i\to\infty} \rho_i = 0$,
    so for all but finitely many $i$,
    we have $\rho_i < \rho$.
    Transitions with $\rho_i < \rho$ 
    cannot be used to produce states in $\Lambda^m_\rho$,
    as their rate constants smaller than the definition of $\Lambda^m_\rho$ allows.
    This shows that $|\Lambda^m_\rho| < \infty$. 
    
    By hypothesis all $s \in \Lambda^0$ satisfy $\vi(s) \geq \alpha n$.
    Fix a particular $s \in \Lambda^0$.
    Let $k = \alpha n$ in Corollary~\ref{cor:chernoff-decay};
    then
    \[
        \Pr{ (\exists t \in [0,1])\ \timeRand{t}{s} < \frac{\alpha n}{81} }
    \leq 
        2^{- \alpha n / 81}.
    \]
    By the union bound,
    \begin{equation}\label{ineq-Lambda0-high}
        \Pr{ (\exists s \in \Lambda^0)(\exists t \in [0,1])\ \timeRand{t}{s} < \frac{\alpha n}{81} }
    \leq
        |\Lambda^0| 
        2^{- \alpha n / 81}.
    \end{equation}
    
    That is, with high probability, 
    all states in $\Lambda^0$ have ``large'' count 
    (at least $\frac{\alpha n}{81}$) for the entire first unit of time.
    Call this event $H(\Lambda^0)$ 
    (i.e., the complement of the event in \eqref{ineq-Lambda0-high}).
    
    We complete the proof by a ``probabilistic induction'' 
    on $i \in \{0,1,\ldots,m\}$ as follows.
    We show a sequence $\delta_0 > \delta_1 > \ldots > \delta_{m} > 0$ such that the following holds.
    Inductively assume that for all $s \in \Lambda^i_\rho$ and all 
    $t \in \left[ \frac{i}{m+1},1 \right]$, 
    $\timeRand{t}{s} \geq \delta_i n$.
    Call this event $H(\Lambda^i_\rho)$.
    Then we will show that
    assuming $H(\Lambda^i_\rho)$ holds, 
    with high probability $H(\Lambda^{i+1}_\rho)$ holds,
    i.e., for all $s \in \Lambda^{i+1}_\rho$ 
    and for all 
    $t \in \left[ \frac{i+1}{m+1}, 1 \right]$,
    $\timeRand{t}{s} \geq \delta_{i+1} n$.
    The base case is established by~\eqref{ineq-Lambda0-high} for 
    $\delta_0 = \frac{\alpha}{81}$.
    
    We use Chernoff bounds for binomial random variables,
    which state that, 
    for $1 \leq i \leq k$,
    if each $\mathbf{X}_i$ is an independent 0/1-random variable with 
    $\Pr{ \mathbf{X}_i = 1 } = p$,
    defining $\mathbf{X} = \sum_{i=1}^k \mathbf{X}_i$
    and $\mu = \E{\mathbf{X}} = kp$,
    then for $0 < \beta \leq 1$,
    $
        \Pr{ \mathbf{X} \leq (1-\beta) \mu } 
    \leq 
        e^{-\mu \beta^2 / 2}
    $
    and
    $
        \Pr{ \mathbf{X} \geq (1+\beta) \mu } 
    \leq 
        e^{-\mu \beta^2 / 3}.
    $
    
    To see that the inductive case holds, 
    fix a particular state $s \in \Lambda^{i+1}_\rho \setminus \Lambda^{i}_\rho$; 
    then $s \in \mathrm{PROD}(\Lambda^{i}_\rho)$.
    By the definition of $\mathrm{PROD}(\Lambda^{i}_\rho)$, 
    $s$ is produced by a transition of the form $x,y \rxn^{\rho'} s,s'$
    where $x,y \in \Lambda^i_\rho$
    and $\rho' \geq \rho$.
    At time $t$,
    the given transition has probability 
    $\rho' \cdot \timeRand{t}{x} \cdot \timeRand{t}{y} / {n \choose 2}$
    (if $x \neq y$)
    or
    $\rho' \cdot (\timeRand{t}{x} \cdot (\timeRand{t}{x} - 1) / 2) / {n \choose 2} $
    (if $x = y$)
    of occurring in the next interaction.
    We make the worst-case assumption that the probability is the latter probability,
    which is smaller when we also make the worst-case assumption 
    $\timeRand{t}{x} = \timeRand{t}{y} = \delta_i n$,
    and substitute $\rho$ for $\rho'$ since $\rho' \geq \rho$.
    
    By the induction hypothesis $H(\Lambda_i)$, 
    for all $t \in \left[ \frac{i}{m+1},1 \right]$, 
    $\timeRand{t}{x} \geq \delta_i n$ and $\timeRand{t}{y} \geq \delta_i n$.
    So for each interaction between time $\frac{i}{m+1}$ and $\frac{i+1}{m+1}$,
    the probability that it executes transition
    $x,y \rxn^{\rho'} s,s'$
    is at least
    \[
        \frac{\rho \delta_i n (\delta_i n -1 ) / 2 }{ {n \choose 2} }
    =
        \frac{\rho \delta_i n (\delta_i n -1 ) / 2 }{ n (n-1) / 2 }
    =
        \frac{\rho \delta_i (\delta_i n - 1)}{n-1}
    >
        \frac{\rho \delta_i (\delta_i n)}{n}
    =
        \rho \delta_i^2.
    \]
    There are $\frac{n}{m+1}$ interactions in that time interval.
    Thus the number of times $x,y \rxn^{\rho'} s,s'$ executes in that interval is stochastically dominated by a binomial random variable $\mathbf{X}^+$,
    with $k = \frac{n}{m+1}$ trials and probability of success $p = \rho \delta_i^2$,
    and $\mu = \E{\mathbf{X}^+} = kp = \frac{n \rho \delta_i^2}{m+1}$.
    By the Chernoff bound,
    setting $\beta = \frac{1}{2}$,
    \[
        \Pr{ \mathbf{X}^+ \leq \frac{ n \rho \delta_i^2 }{ 2(m+1) } }
    =
        \Pr{ \mathbf{X}^+ \leq (1 - \beta) \mu } 
    \leq
        e^{- \mu \beta^2 / 2}
    =
        \exp{ - \frac{n \rho \delta_i^2}{8(m+1)} }.
    \]
    
    The above analysis lower bounds how many times transition $x,y \rxn^{\rho'} s,s'$ executes,
    producing $s$ each time.
    We now upper bound how many times $s$ is consumed in this same interval.
    We are trying to show that $s$ gets to a large count,
    so we make the worst-case assumption that its count starts at 0.
    Any any time,
    out of $\binom{n}{2}$ pairs of agents,
    at most $s(n-1)$ of those pairs have at least one agent in state $s$.
    So at time $t$,
    the probability that the next transition consumes $s$ is at most 
    $
        \frac{s(n-1)}{\binom{n}{2}}
    =
        2 \frac{\timeRand{t}{s}}{n}.
    $
    
    Prior to $s$ reaching count 
    $n \rho \delta_i^2 / 32$,
    we can make the worst case assumption that the probability of each transition consuming $s$ is exactly
    $2 \frac{n \rho \delta_i^2 / 32}{n} 
    = \rho \delta_i^2 / 16.$
    In this worst case the number of transitions consuming $s$ in the $k = n/m$ interactions in the time interval 
    $\left[ \frac{i}{m+1}, \frac{i+1}{m+1} \right]$
    is stochastically dominated by $2 \mathbf{X}^-$,
    where $\mathbf{X}^-$ is a binomial random variable
    with
    $k = n/(m+1)$ trials and probability of success $p = \rho \delta_i^2 / 16$.
    (We consider $2\mathbf{X}^-$ instead of $\mathbf{X}^-$ to account for the fact that each transition in the worst case could consume 2 copies of $s$.)
    Apply the Chernoff bound
    with $\mu = \E{\mathbf{X}^-} = kp = \frac{n \rho \delta_i^2}{16(m+1)}$
    and $\beta = 1$
    to give
    \begin{eqnarray*}
        \Pr{ \mathbf{X}^- \geq \frac{ n \rho \delta_i^2 }{ 8(m+1) } }
    &=&
        \Pr{ \mathbf{X}^- \geq (1+\beta) \mu }
    \leq 
        e^{- \mu \beta^2 / 3}
    = 
        \exp{- \frac{ n \rho \delta_i^2 }{ 48(m+1) } }.
    \end{eqnarray*}
    
    Thus
    \[
        \Pr{ 2 \mathbf{X}^- \geq \frac{ n \rho \delta_i^2 }{ 4(m+1) } }
    \leq
        \exp{- \frac{ n \rho \delta_i^2 }{ 48(m+1) } }.
    \]
    Note that this count threshold is half the count threshold we derived for the lower bound on the number of transitions $x,y \rxn^{\rho} s,s'$ producing $s$.
    Thus,
    applying the union bound to these two events to bound $\mathbf{X}^+ - 2\mathbf{X}^-$,
    the \emph{net} production of $s$ 
    (number produced minus number consumed),
    we have that
    \begin{eqnarray*}
        \Pr{ \mathbf{X}^+ - 2\mathbf{X}^- \leq \frac{ n \rho \delta_i^2 }{ 4(m+1) } }
    &\leq&
        \Pr{ \mathbf{X}^+ \leq \frac{ n \rho \delta_i^2 }{ 2(m+1) } 
            \text{ or }
             2 \mathbf{X}^- \geq \frac{ n \rho \delta_i^2 }{ 4(m+1) }
        }
    \\&\leq&
        \exp{- \frac{ n \rho \delta_i^2 }{ 8(m+1) } }
    +
        \exp{- \frac{ n \rho \delta_i^2 }{ 48(m+1) } }
    \\&<&
        2 \cdot \exp{- \frac{ n \rho \delta_i^2 }{ 48(m+1) } }.
    \end{eqnarray*}
    So with probability at least 
    $
        1 - 2 \cdot \exp{- \frac{ n \rho \delta_i^2 }{ 48(m+1) } },
    $
    at least
    $
        n \rho \delta_i^2 / (4(m+1))
    $
    net copies of $s$ are produced at some point in the time interval 
    $\left[ \frac{i}{m+1}, \frac{i+1}{m+1} \right]$.
    
    Letting $k = n \rho \delta_i^2 / (4 (m+1))$ in Corollary~\ref{cor:chernoff-decay},
    with probability at least
    $1 - 2^{- k / 81} = 1 - 2^{- n \rho \delta_i^2 / (324 (m+1))},$
    we have that $\timeRand{t}{s} \geq \delta_i n / 81$
    for all times 
    $t \in \left[ \frac{i+1}{m+1}, 1 \right]$.
    Setting
    $\delta_{i+1} = (k/n) / 81 = \rho \delta_i^2 / (324 (m+1))$
    proves the inductive case with probability of failure at most
    \[
        2 \cdot \exp{- \frac{ n \rho \delta_i^2 }{ 48(m+1) } }
    +
        2^{ - \delta_{i+1} n }
    \]
    
    By the union bound over all $|\Lambda^m_\rho|$ states in all levels of induction,
    setting $\delta = \delta_m = \rho^m (\alpha/2)^{2^m} / ( 324 (m+1) )^m$,
    noting that $\delta \leq \delta_i$ for all $0 \leq i \leq m$,
    with probability most
    \[
        |\Lambda^m_\rho|
        \left( 
            2 \cdot
            \exp{- \frac{ n \rho \delta^2 }{ 48(m+1) } }
        +
            2^{ - \delta n }
        \right),
    \]
    the count of all states in $\Lambda^m_{\rho}$
    fails to reach at least 
    $\delta n$ by time $t = 1$.
    By setting $n_0$ sufficiently large,
    the above probability is $< 1$ for all $n = n_0$
    (and therefore for all greater $n$ as well).
    By setting $\epsilon > 0$ sufficiently small,
    this probability is at most $2^{- \epsilon n}$.
}

\opt{sub,final}{
    \opt{sub}{A self-contained proof is in Section~\ref{app:termination}.}
    \opt{final}{A self-contained proof is in~\cite{doty2018efficientArxiv}.}
    Intuitively,
    Lemma~\ref{lem:timer} can be used to prove Theorem~\ref{thm:main-termination} in the following way.
    In some ``small'' population size $n_0$,
    the terminal signal appears.
    The set of states $\Lambda' \subseteq \Lambda$ appearing with the terminal signal is constant size.
    Lemma~\ref{lem:timer} states that for any constant-size $\Lambda' \subseteq \Lambda$,
    in all sufficiently large population sizes,
    all states in $\Lambda'$ appear in constant time with high probability,
    so the termination signal appears prematurely in larger populations.
    This is fairly straightforward for deterministic transition functions,
    but it requires some care to handle a randomized protocol.
}

\opt{full,final}{
    We now use Lemma~\ref{lem:timer} to formally prove Theorem~\ref{thm:main-termination}.

    \begin{proof}[Proof of Theorem~\ref{thm:main-termination}]
        Assume $P$ is $\kappa$-$t$-terminating;
        we will show $t(n) = O(1)$.
        Let $(\vc_i)_{i=1}^\infty$ be an infinite sequence of $\alpha$-dense initial configurations in $I$.
        Dickson's Lemma~\cite{dicksonslemma} states that every infinite sequence in $\N^k$ has an infinite nondecreasing subsequence,
        so assume without loss of generality that $\vc_i \leq \vc_{i+1}$ for all $i \in \N$.
        Let $\Lambda^0 = \{ s \in \Lambda \mid \vc_0(s) > 0 \}$ be the set of states present in $\vc_0$.
        
        By hypothesis $\Pr{P \text{ terminates from } \vc_0} \geq \kappa > 0$.
        Thus there is at least one 
        finite execution $\calE$
        starting with $\vc_0$ and ending in a terminated configuration.
        Let $m = |\calE|$ be the length of this execution.
        Let $\rho$ be the minimum rate constant of any transition in $\calE$.
        Then every state appearing in configurations in 
        $\calE$
        is $m$-$\rho$-producible from $\vc_0$,
        i.e., is in $\Lambda^m_\rho$ where $\Lambda^0 = \{ s \in \Lambda \mid \vc_0(s) > 0\}$
        is the set of states present in $\vc_0$.
        
        For any $\ell \geq 1$,
        since $\vc_0 \leq \vc_\ell$,
        all states in $\Lambda^m_\rho$ are $m$-$\rho$-producible from $\vc_\ell$ as well.
        By Lemma~\ref{lem:timer},
        there are constants $\epsilon,\delta,n_0 > 0$ such that, 
        for all $\ell \in \N$ 
        such that $n = \|\vc_\ell\| \geq n_0$,
        letting $\timeRand{1}{s}$ be the random variable denoting the count of $s$ at time $1$,
        assuming at time $0$ the configuration is $\vc_{\ell}$,
        $
            \Pr{ \left( \forall s \in \Lambda^m_\rho \right)\ \timeRand{1}{s} \geq \delta n} 
            \geq 1 - 2^{-\epsilon n}.
        $
        
        However, $\Lambda^m_\rho$ contains terminated states,
        so for all $\vc_{\ell}$ with $\|\vc_\ell\| \geq n_0$,
        with probability $\geq 1 - 2^{-\epsilon n}$,
        $P$ terminates within time $1$.
        Since $2^{-\epsilon n} < \kappa$ for sufficiently large $n$,
        this implies that 
        if $P$ is $\kappa$-$t$-terminating,
        then $t(n) \leq 1$ for sufficiently large $n$.
        Thus $t(n) = O(1)$.
    \end{proof}
}

\opt{final,full}{
    Observe how the assumption of uniformity is used in the proof:
    we take a set of transitions used on the population $\vc_0$ 
    and apply it to a larger population $\vc_\ell$.
    In a nonuniform protocol,
    the transitions may not be legal to apply to $\vc_\ell$.
    As a concrete example, 
    in a nonuniform protocol, 
    an agent increments a counter
    using transitions such as $c_7,x \to c_8,y$
    until the counter exceeds $\log n$,
    then produces a termination signal $t$
    via a transition $c_8,x \to t,y$.
    The transition $c_8,x \to t,y$ producing this signal is not legal in a population larger than twice $n$,
    since the value $\log n$ is at least 1 larger in such a protocol.
    In this example,
    the transition of the larger protocol with the same input states 
    simply increments the counter without producing a termination signal:
    $c_8,x \to c_9,y$.
}

\vspace{0.3cm}
\noindent
{\bf Acknowledgements.}
We are grateful to Eric Severson for helpful comments
and anonymous reviewers for their suggestions,
which vastly improved the paper.
The second author thanks James Aspnes for discussions that stimulated a key idea used in the main protocol.

\newpage
\appendix



\newpage
\section{Proofs for correctness of size estimation protocol}
\label{app:protocol}

This section contains proofs of lemmas required to analyze the correctness and time/space complexity of the size estimation protocol of Theorem~\ref{thm:main-size-estimation-protocol}.

\restateLemboundOnCardinalityOfA
\proofLemboundOnCardinalityOfA

\begin{lemma}[\cite{AAE08}]\label{lem:epidemic}
    Let $T$ denote the time to complete an epidemic.
    Then
    $\E{T} = \frac{n-1}{n} H_{n-1}$,
    $\Pr{T < \frac{1}{4} \ln n} < 2 e^{-\sqrt{n}}$,
    and for any $\alpha_u > 0$,
    $\Pr{T > \alpha_u \ln n} < 4 n^{- \alpha_u/4+1}$.
\end{lemma}

\restateCorPartialEpidemic
\proofCorPartialEpidemic

\begin{corollary}\label{cor:grvalue}
    The \gr\ value
    is in the interval of $[\log n - \log \ln n-2, 2\log n-1]$ with probability at least $1-1/n-e^{-n/18}$.
\end{corollary}

\begin{corollary}\label{cor:boundONtime}
    The number of interactions in each epoch in the $\sizest$ is in the interval 
    $[\cte \log n - \cte \log \ln n, 191\log n]$ with probability $\geq 1-1/n-e^{-n/18}$. 
\end{corollary}

\begin{proof}
    By Corollary~\ref{cor:exactboundonepoch}, agents should count up to $65\ln n \leq 94 \log n$ before moving to the next epoch. if we set the threshold of the \counter\ to $\cte \cdot \clock$, $\cte \log n - \cte  \log \ln n \geq 94 \log n$ then the \counter\ variable will be in the interval of $[\cte \log n - \cte \log \ln n, 190\log n+\cte]$ with high probability ($190 \log n + \cte \leq 191 \log n$ for $n \geq 2$).
\end{proof}

\begin{corollary}\label{cor:boundONcoefficient}
    The number of epochs in the $\sizest$ is in the interval 
    $[\constantNumGrv\log n - \constantNumGrv\log \ln n, 11\log n]$ with probability $\geq 1-1/n-e^{-n/18}$.
\end{corollary}

\begin{proof}
    By Corollary~\ref{cor:sumofgeomConcreteK}, to achieve the additive error of \logSizeError\ for our protocol the number of geometric random variables should be $\geq 4\log n$. By setting the threshold of the number of phases to $\constantNumGrv \times \clock$, for $n\geq 200$, $\constantNumGrv \log n - \constantNumGrv \log \ln n \geq 4\log n$. The number of phases will be in the interval of $[\constantNumGrv\log n - \constantNumGrv\log \ln n, 10\log n + \constantNumGrv]$ with high probability ($10 \log n + \constantNumGrv\leq 11 \log n$ for $n\geq 2$).
\end{proof}

\opt{sub,final}{
\restateCorSumOfGeometricAmongAagents
\proofCorSumOfGeometricAmongAagents

\restateCorTimeComplexity
\proofCorTimeComplexity

\restateLemBoundOnepoch
\proofLemBoundOnepoch

\restateLemSpace
\proofLemSpace
}

\restateTHMmainSizeEstimationProtocol
\proofTHMmainSizeEstimationProtocol

\newpage
\section{Size estimation with no access to random bits}
\label{app:protocol-det}

Our protocol uses uniform random bits in multiple places. 
In this section we use the inherent randomness of the uniform random scheduler to simulate our own random bits similar to that of~\cite{sudo2018logarithmic}. 
In the protocol, agents start by dividing in two groups of \f\ and \a. \a\ agents are responsible to compute the algorithm while the \f\ agents only provide fair coin flips. 
When an \a\ agent interact with an \f\ agent, 
with probability exactly $\frac{1}{2}$ each can be the sender or the receiver;
this is used to assign the random bit to the $\a$ agent.

Agents initially have no role ($\x$),
and partition into roles via $\x,\x \to \a,\f$.
Since this takes $\Theta(n)$ time to complete,
we add transitions $\a,\x \to \a,\f$ and $\f,\x \to \f,\a$,
converging in $O(\log n)$ time, 
with the price of deviating from $\frac{n}{2}$ for each role. 
By Lemma~\ref{lem:boundOnCardinalityOfA} this deviation is $O(\sqrt{n \ln n})$,
increasing the size estimation error by merely a constant additive factor.\footnote{
    This mechanism of splitting the population approximately in two works for our protocol,
    because the number of $\a$ agents is likely to be so close to $n/2$ that our estimate of $\log n$ is reduced by an additive factor likely to be very close to $-1$.
    If some downstream protocol requires all agents to participate in the algorithm
    (e.g., for predicate computation),
    then a similar but more complex scheme works instead:
    All agents count their number of interactions mod 2,
    acting in the $\a$ role on even interactions and the $\f$ role on odd interactions.
    This implies a similar constant-factor slowdown,
    and it obtains the required independence of coin flips from each other and from algorithm steps.
}

All agents start at $\phase =0$. 
The $\a$ agent use random bits,
obtained by a synthetic coin technique 
(due to~\cite{sudo2018logarithmic}) 
through interaction with $\f$ agents,
to generate a geometric random variable. 
In protocol~\ref{subprotocol:genClk} an \a\ agent starts with $1$ in its \clock\ field and increases it as long as it is the sender (``coin flip = tails'') 
in an interaction with an \f\ agent. 
When an \a\ agent interacts as the receiver (``coin flip = heads'') with an \f\ agent, 
it will set a flag meaning that generating the \clock\ variable is completed.
Those agents who completed generating their \clock\ value will start propagating the maximum one they have generated. Since we use this \clock\ value for all early estimation of $\log n$, each time an agent finds out there was a greater value for the \clock\ than its own, it will reset all other computations that might have happened.

By Lemma~\ref{lem:clkvalue}, the maximum \clock\ amongst the population is a $2$ factor estimation of $\log n$. 
When any agent updates its $\clock$ with a new maximum,
it restarts the entire downstream protocol via $\reset$.
Once the maximum $\clock$ value is generated in the population,
it propagates (triggering $\reset$) by epidemic in $O(\log n)$ time.
The \clock\ variable could be used to estimate $K$, which is the number of independent additional geometric random variables each agent will generate. We also use \clock\ to set the leaderless phase clock inside each agent. At each epoch, agents will generate one new geometric random variable and propagate its maximum. All $\a$ agents counts their number of interactions in their \counter\ variable. If any agent's \counter\ reaches $\cte.\clock$, their \phase\ variable increases by one and set they $\counter =0$.

In the \sizest\ protocol, all agents in role \a\ will finally generate $K =\constantNumGrv \cdot \clock$ geometric random variable and store a sum of maximum one generated for each phase. 
For each one of the geometric random variables, agents start with $1$ and increase it as long as they interact as the sender with \f\ agents. Similar to generating the \clock\ variable, whenever, an \a\ agent interact as the receiver with an \f\ agent, generating a geometric random variable is completed and they will move on to propagate the maximum they have generated. 
Note that 
those agents who completed generating their \gr\ value will start propagating the maximum.
Once all agents reach $\phase =\constantNumGrv \cdot \clock$ they set $\doneEST = \true$ and $\outpt = \frac{\ksum}{\phase}+1$.
We use $|\a|$, $|\f|$ for the cardinality of \a\ and \f\ agents respectively.

\begin{algorithm}[ht]
	\floatname{algorithm}{Protocol}
	\caption{$\sizest(\rec, \sen)$}
	\label{protocol:estimation6}
	\begin{algorithmic}[100]		
        
		\LeftComment {initial state of agent:
		
		    $\role = \x,$
		    
    		$\counter = 0,		
    		\ksum = 0,		
    		\phase = 0,$
    		
    		$\gr = 1,
    		\clock = 1,$
		
		    $\doneC = \false, $
		    
		    $\doneG = \false, $
		    
		    $\doneEST = \false$
		}
		
    	\State{$\partition(\rec, \sen)$}
    	\If {$\rec.\role = \a $} 
    		\State{$\rec.\counter \gets \rec.\counter +1$}
		    \State{$\timerDone(\rec)$}
        \EndIf
        \If {$\sen.\role = \a $} 
    		\State{$\sen.\counter \gets \sen.\counter +1$}
    		\State{$\timerDone(\sen)$}
        \EndIf
        
        \If{\text{one agent is in state \f\ and one is in state \a}}
            \State{$\agentA \gets $ the agent in state \a}
            \If{\agentA.\doneC\ = \false}
		        \State{$\genClk(\rec, \sen)$}
		    \ElsIf{\agentA.\doneG\ = \false}
                \State{$\genGr(\rec, \sen)$}
            \EndIf
        \EndIf
        \If{\text{both agents have $\role = \a$ and $\doneG = \true$}}
            \State{$\epidemicClk(\rec, \sen)$}
		    \If{\text{both agents have $\doneG = \true$}}
		    \State{$\epidemicPhase(\rec,\sen)$}
		    \State{$\epidemicGr(\rec, \sen)$}
		    \EndIf
		    
		    \If{\sen.\doneEST}	
		        \State{$\outpt \gets \frac{\ksum}{\phase}+1
		        $}
	        \EndIf
		\EndIf
	
	\end{algorithmic}
\end{algorithm}
\begin{algorithm}[!htbp]
	\floatname{algorithm}{Subprotocol}
	\caption{$\partition(\rec, \sen)$}
	\label{subprotocol:partition6}
	\begin{algorithmic}[100]
	    \LeftComment{Partition the population in two almost equal size subpopulations.}
	    
        \If {$\sen.\role = \x , \rec.\role = \x$} 
            \State{$\sen.\role \gets \a$}
		    \State{$\rec.\role \gets \f$}
		\ElsIf{$\sen.\role = \a , \rec.\role = \x$}
		    \State{$\rec.\role \gets \f$}
		\ElsIf{$\sen.\role = \f , \rec.\role = \x$}
		    \State{$\rec.\role \gets \a$}
		\EndIf
		
	\end{algorithmic}
\end{algorithm}

\begin{algorithm}[ht]
	\floatname{algorithm}{Subprotocol}
	\caption{$\genClk(\rec, \sen)$}
	\label{subprotocol:genClk}
	\begin{algorithmic}[100]
	    \LeftComment{Generate one geometric random variable for \clock.}
        \If {$\sen.\role=\a$} \Comment{This is only called if exactly one agent has \role=\a.}
            \State{$\sen.\clock \gets \sen.\clock +1$}
        \ElsIf{$\rec.\role = \a$}
            \State{$\rec.\doneC \gets \true$}
            \State{$\rec.\clock \gets \rec.\clock+2$}
		\EndIf
	\end{algorithmic}
\end{algorithm}

\begin{algorithm}[ht]
	\floatname{algorithm}{Subprotocol}
	\caption{$\epidemicClk(\agent1, \agent2)$}
	\label{subprotocol:epidemic_clk6}
	\begin{algorithmic}[100]
	    \LeftComment{Maximum generated geometric variable for $\clock$ will be propagated.}
        \If {$\agent1.\clock < \agent2.\clock$} 
            \State{$\agent1.\clock \gets \agent2.\clock$}
		    \State{$\reset(\agent1)$}
		\ElsIf{$\agent2.\clock < \agent1.\clock$} 
            \State{$\agent2.\clock \gets \agent1.\clock$}
		    \State{$\reset(\agent2)$}
		\EndIf
	\end{algorithmic}
\end{algorithm}

\begin{algorithm}[!htbp]
 	\floatname{algorithm}{Subprotocol}
 	\caption{$\reset(\agent)$}
 	\label{subprotocol:reset6}
 	\begin{algorithmic}[100]
 	\State {$\counter \gets 0$,		
	$\ksum \gets 0$,		
	$\phase \gets 0$,		
	$\gr \gets 1$}
	\State {$\doneG \gets \false$,
	$\doneEST \gets \false$}
 	    
 	     
 	      
		
 		
 	\end{algorithmic}
 \end{algorithm}
 
\begin{algorithm}[!htbp]
	\floatname{algorithm}{Subprotocol}
	\caption{$\genGr(\rec, \sen)$}
	\label{subprotocol:genGr}
	\begin{algorithmic}[100]
	    \LeftComment{Generating one geometric random variable for \gr.}
        \If {$\sen.\role=\a$} \Comment{This is only called if exactly one agent has \role=\a.}
            \State{$\sen.\gr \gets \sen.\gr +1$}
        \ElsIf{$\rec.\role = \a$}
            \State{$\rec.\doneG \gets \true$}
		\EndIf
	\end{algorithmic}
\end{algorithm}
 
\begin{algorithm}[!htbp]
	\floatname{algorithm}{Subprotocol}
	\caption{$\epidemicGr(\agent1, \agent2)$}
	\label{subprotocol:epidemic_gr6}
	\begin{algorithmic}[100]
	    \LeftComment{Maximum generated geometric variable for $\gr$ will be propagated.}
	    \If{$\agent1.\phase = \agent2.\phase$}
            \If {$\agent1.\gr < \agent2.\gr$} 
                \State{$\agent1.\gr \gets \agent2.\gr$}
    		\ElsIf{$\agent2.\gr < \agent1.\gr$} 
                \State{$\agent2.\gr \gets \agent1.\gr$}
    		\EndIf
		\EndIf
	\end{algorithmic}
\end{algorithm}

\begin{algorithm}[ht]
	\floatname{algorithm}{Subprotocol}
	\caption{$\timerDone(\agent)$}
	\label{subprotocol:timer_done6}
	\begin{algorithmic}[100]
	    \LeftComment{Agents compare their \counter\ value to the specified threshold.}
        \If {$\agent.\counter = \cte\times \agent.\clock$, $\agent.\doneEST=\false$} 
            \State{$\agent.\phase \gets \agent.\phase +1$}
            \State{$\updateSum(\agent)$}
        \EndIf
		\If{$\agent.\phase = \constantNumGrv \times \agent.\clock$} 
            \State{$\agent.\doneEST \gets \true$}
		\EndIf
	\end{algorithmic}
\end{algorithm}

\begin{algorithm}[ht]
	\floatname{algorithm}{Subprotocol}
	\caption{$\epidemicPhase(\agent1, \agent2)$}
	\label{subprotocol:epidemic_phase6}
	\begin{algorithmic}[100]
	    \LeftComment{The maximum \phase\ will be propagated.}
        \If {$\agent1.\phase < \agent2.\phase$} 
            \State{$\agent1.\phase \gets \agent2.\phase$}
            \State{$\updateSum(\agent1)$}
        \ElsIf {$\agent2.\phase < \agent1.\phase$} 
            \State{$\agent2.\phase \gets \agent1.\phase$}
            \State{$\updateSum(\agent2)$}
        \EndIf
	\end{algorithmic}
\end{algorithm}
 
\begin{algorithm}[!htbp]
	\floatname{algorithm}{Subprotocol}
	\caption{$\updateSum(\agent)$}
	\label{subprotocol:update_sum6}
	\begin{algorithmic}[100]
            \LeftComment{The \agent\ accumulates its current value of \gr\ in \ksum\ and move to the next \phase:}
		    
		    \State{$\agent.\ksum \gets \agent.\ksum +\agent.\gr$}
		    
            \State{$\agent.\counter \gets 0$, $\agent.\gr \gets 1$}
            
            \State{$\agent.\doneG \gets \false$}
            
	\end{algorithmic}
\end{algorithm}



\opt{sub}{The following corollary shows that close to half of agents end up in role $\a$. A generalized Lemma and its proof appears in the appendix.
}

The following lemma shows that generating one geometric random variable takes $O(\ln n)$ with high probability.
\opt{sub}{A generalized Lemma with a proof appears in the appendix.}
\opt{full}{
\begin{lemma}\label{lem:timeToGenerateGR}
    Let $\alpha>0$.
    With probability $\geq 1 - (3/n)^{\alpha-1} - 2 e^{-n/18}$,
    each of protocols \genClk\ and \genGr\ require at most $4\alpha \ln n$ time 
    for all agents to generate one geometric random variable.
\end{lemma}
\begin{proof}
    In this proof, we use the facts,
    following from Corollary~\ref{cor:lazyboundOnCardinalityOfA},
    that $|\a| \geq n/3$ with probability $\geq 1 - e^{-n/18}$,
    and similarly for $|\f| \geq n/3$.
    We model time for generating all geometric random variables as the time to collect all $|\a|$ coupons in a modified coupon problem.
    In the modified problem,
    after $i-1$ coupons have been collected,
    the probability of collecting the $i$'th coupon on the next try,
    rather than being $p_i = \frac{1}{i-1}$,
    is instead 
    $p_i 
    = \frac{|\a-(i-1)| \cdot |\f| \cdot 1/2}{\binom{n}{2}}
    = \frac{|\a-(i-1)| \cdot |\f|}{n(n-1)}$.
    This models that the two agents must be in roles $\a$,$\f$, respectively,
    and the agent in role $\a$ must be the receiver, 
    to complete the generating of the $\a$ agent's geometric random variable.
    Let $\mathbf{T}$ be the time to collect all $|\a|$ coupons, and let $\mathbf{t}_i$ be the time to collect the $i$'th coupon after $i-1$ coupons have been collected,
    with $\E{\mathbf{t}_i} = 1 / p_i$. 
    Then
    \begin{eqnarray*}
            \E{\mathbf{T}}
        &=&
            \sum_{i=1}^{|\a|} \E{\mathbf{t}_i} 
        =
            \sum_{i=1}^{|\a|} \frac{1}{p_i}
        =
            \sum_{i=1}^{|\a|} \frac{n(n-1)}{|\f|\cdot|\a-(i-1)|}
        \\&=&           \frac{n(n-1)}{|\f|}\sum_{i=1}^{|\a|} \frac{1}{i}
        \leq 
            \frac{n(n-1) H_{|\a|}}{|\f|}
        \\&\leq&
            \frac{n(n-1) H_{|\a|}}{\frac{n}{3}} 
            \ \ \ \ \ \text{since $|\f| \geq \frac{n}{3}$ by Corollary~\ref{cor:lazyboundOnCardinalityOfA}}
        \\&=&
            3 (n-1) H_{|\a|}.
    \end{eqnarray*}
    For the upper bound, 
    fix some coupon that has not been collected 
    ($\a$ agent that has not completed its geometric random variable).
    The probability that the next interaction collects \emph{that} coupon is $\frac{|\f|}{2 \binom{n}{2}}$.
    If $T_m = 3 \alpha (n-1)\ln{|\a|}$,
    then the probability of that not getting selected after $T_m$ time is then
    $= \left(1- \frac{|\f|}{2\cdot \binom{n}{2}}\right)^{T_m}\leq \left(1- \frac{n}{3n(n-1)}\right)^{T_m}$. 
    By the union bound over all agents in role \a, the probability that there exists a coupon that is not selected after $T_m$ time is 
    $\leq |\a| \left(1- \frac{1/3}{n-1} \right)^{\alpha \cdot 3(n-1)\ln{|\a|}} 
    \leq |\a| e^{- \alpha \ln {|\a|} } 
    = \frac{1}{|\a|^{\alpha-1}} 
    \leq \left( \frac{3}{n} \right)^{\alpha-1}$,
    where the last inequality follows from $|\a| \geq n/3$ with probability $\geq 1 - e^{-n/18}$.
\end{proof}
}
Setting $\alpha =3$ in Lemma~\ref{lem:timeToGenerateGR} gives the following.

\begin{corollary}\label{cor:exacttimeToGenerateGR}
Protocols \genClk\ and \genGr\ require at most $12 \ln n$ time to generate one geometric random variable with probability $\geq 1 - \frac{9}{n^{2}}- 2e^{-n/18}$.
\end{corollary}

In each epoch, one geometric random variable 
(in the first epoch \clock\ and in the subsequent epochs \gr) 
is generated and its maximum will be propagated by epidemic among the population. We set the time of each epoch equal to the required time of generating one plus the time for completion of an epidemic.
To analyze the time complexity of our protocol, 
we require the time bounds for completing an epidemic from the paper~\cite{AAE08}. 
The current form is taken from~\cite{doty2018exact}.
For all $n \in \N^+$,
let $H_n = \sum_{k=1}^n \frac{1}{k}$ denote the $n$'th harmonic number.
Note that $\ln n \leq \frac{n-1}{n} H_{n-1} \leq 1 + \ln n$.

\opt{final}{
\begin{lemma}[\cite{AAE08}]\label{lem:epidemic}
Let $T$ denote the time to complete an epidemic.
Then
$\E{T} = \frac{n-1}{n} H_{n-1}$,
$\Pr{T < \frac{1}{4} \ln n} < 2 e^{-\sqrt{n}}$,
and for any $\alpha_u > 0$,
$\Pr{T > \alpha_u \ln n} < 4 n^{- \alpha_u/4+1}$.
\end{lemma}
}

The following corollary describes an epidemic in a subpopulation.
This refers to some subset $S$ of the population executing epidemic transitions only among themselves,
which slows down the epidemic by only a constant factor if $|S| = \Omega(n)$.

By Corollary~\ref{cor:exacttimeToGenerateGR}, an agent generate a geometric random variable after $12\ln n$ time, and by Corollary~\ref{cor:partialEpidemic-c-equals-3}, $24\ln n$ time is sufficient to propagate the maximum generated one w.h.p. Thus, we can obtain the following corollary.

\begin{corollary}\label{cor:generateGRplusEpidemic6}
    Suppose at lease $n/3$ agents are in role \a.
    Let $\mathbf{T}$ be the time for them generate one geometric random variable 
    and propagate its maximum to the whole subpopulation of $\a$'s.
    Then $\Pr{\mathbf{T} > 36 \ln n} < 10n^{-2}$.
\end{corollary}

The next lemma bounds the number of interactions an agent has in a given time,
and it is the basis of the leaderless phase clock we use.
\opt{sub}{It is proven in the appendix.}
\opt{final}{It is proven in~\cite{doty2018efficientArxiv}.}
It follows from a simple Chernoff bound on the number of interactions involving a single agent in a given window of time.

\begin{corollary}\label{cor:exactboundonepoch6}
Each agent has $\geq 96 \ln n$ interactions
in time $36 \ln n$
with probability $\leq 1/n$.
\end{corollary}

By Lemma~\ref{lem:boundONepoch} each agent has at most 
$\left(2\cdot 36 +\sqrt{12 \cdot 36}\right) \ln n \leq 96 \ln n \leq 139 \log n$ interactions in the time that it takes to generate and propagate maximum of one geometric random variable. 
Thus, each agent should count up to $139 \log n$ for its leaderless phase clock,
to ensure that with high probability none reaches that count until the maximum geometric random variable is known to all agents. 
However, agents are not aware of any approximation of $\log n$. In the \sizest\ protocol, agents use their \clock\ variable for this approximation. 
As mentioned, all the agents in role \a\ start by generating one geometric random variable \clock.
The maximum in the population is used as a weak (constant factor) approximation of $\log n$. 
\opt{full}{
Corollary~\ref{cor:half-geom-tighter-bounds} says that the maximum of $|\a|$ geometric random variables is in the interval of
$[\log |\a| - \log \ln |\a|, 2\log |\a|]$ 
with probability at least $1-1/|\a|$. However, we are using the \clock\ and \gr\ variables as an approximation of $\log n$ rather than $\log |\a|$. Lemma~\ref{lem:clkvalue} will give us a bound over the \clock\ value with respect to $\log n$. Corollaries~\ref{cor:grvalue},~\ref{cor:boundONtime},~\ref{cor:boundONcoefficient} use this lemma for a bound over \gr, \counter, and \phase\ values.
}
\opt{sub}{Their statement and proofs appear in the appendix.}
\opt{final}{Their statement and proofs appear in~\cite{doty2018efficientArxiv}.}
    


\opt{final}{
    \begin{corollary}\label{cor:grvalue}
    The \gr\ value generated by \genGr\ is in the interval of $[\log n - \log \ln n-2, 2\log n-1]$ with probability at least $1-1/n-e^{-n/18}$.
    \end{corollary}
    
    \begin{corollary}\label{cor:boundONtime}
    The number of interactions in each epoch in the $\sizest$ is in the interval 
    $[\cte \log n - \cte \log \ln n, 281\log n]$ with probability $\geq 1-1/n-e^{-n/18}$. \end{corollary}
    
    \begin{proof}
    By Corollary~\ref{cor:exactboundonepoch}, agents should count up to $96\ln \leq 139 \log n$ before moving to the next epoch. if we set the threshold of the \counter\ to $\cte \cdot \clock$, $\cte \log n - \cte  \log \ln n \geq 139 \log n$ then the \counter\ variable will be in the interval of $[\cte \log n - \cte \log \ln n, 280\log n+\cte]$ with high probability ($280 \log n + \cte \leq 281 \log n$ for $n \geq 2$).
    \end{proof}
    
    \begin{corollary}\label{cor:boundONcoefficient}
    The number of epochs in the $\sizest$ is in the interval 
    $[\constantNumGrv\log n - \constantNumGrv\log \ln n, 11\log n]$ with probability $\geq 1-1/n-e^{-n/18}$. \end{corollary}
    
    \begin{proof}
    By Lemma~\ref{lem:sumofgeomConcreteK}, to achieve the additive error of \logSizeError\ for our protocol the number of geometric random variables should be $\geq 4\log n$. By setting the threshold of the number of phases to $\constantNumGrv \times \clock$, for $n\geq 200$, $\constantNumGrv \log n - \constantNumGrv \log \ln n \geq 4\log n$. The number of phases will be in the interval of $[\constantNumGrv\log n - \constantNumGrv\log \ln n, 10\log n + \constantNumGrv]$ with high probability ($10 \log n + \constantNumGrv\leq 11 \log n$ for $n\geq 2$).
    \end{proof}
}

The next Lemma bounds the space complexity of our main protocol
by counting the likely range taken by the variables in $\sizest$.
\opt{sub,final}{
    It is proven 
    \opt{sub}{in the Appendix.}
    \opt{final}{in~\cite{doty2018efficientArxiv}.}
}

\restateableLemma{lem:spacesix}{LemSpaceSix}{
    $\sizest$ uses $O(\log^6 n)$ states with probability $\geq 1-O(\log n)/n$.
}{
    With probability at least $1 - O(1/n)$
    (see individual lemma statements for constants in the $O$),
    the set of values possibly taken on by each field are given as follows:
    \begin{center}
    \begin{tabular}{ l l l }
     \clock & $\{1, \ldots, 2\log n+1\}$ 
            & Lemma~\ref{lem:clkvalue} \\ 
     \gr & $\{1, \ldots, 2\log n\}$ 
            & Corollary~\ref{cor:grvalue} \\ 
     \phase & $\{0,\ldots, 11 \log n\}$ 
            & Corollary~\ref{cor:boundONcoefficient} \\
     \counter & $\{0, \ldots, 281 \log n\}$ 
            & Corollary~\ref{cor:boundONtime} \\
     \ksum & $\{0, \ldots, 22 \log ^2 n\}$ 
            & Corollaries~\ref{cor:grvalue},~\ref{cor:boundONcoefficient}
    \end{tabular}
    \end{center}
    
    After each agent sets $\doneEST =\true$, 
    it no longer needs to store the value in \gr\ and can use that space to store the result of $\ksum / \phase +1$ as the output. The probability that each geometric random variable is greater than $2\log n$ is less than $1/n$, by the union bound the probability that any of them is greater than $2\log n$ is less than $\frac{11\log n}{n}$.
}

\opt{full}{\proofLemSpaceSix}

The next corollary bounds the time complexity of protocol \sizest;
the main component of the time complexity is that $\Theta(\log n)$ geometric random variables must be generated and propagated by epidemic among the population,
each epidemic taking $\Theta(\log n)$ time.
\opt{sub}{A proof appears in the appendix.}
\opt{final}{A proof appears in~\cite{doty2018efficientArxiv}.}
\restateableCorollary{cor:timecomplexitysix}{CorTimeComplexitysix}{
    The \sizest\ protocol take $O(\log^2 n)$ time with probability at least $1-1/n$ for all agents set $\doneEST =\true$.
}{
    By Corollary~\ref{cor:generateGRplusEpidemic6}, with probability $\geq 1-(10/n^2)$ generating and propagating the maximum of the \clock\ variable takes at most $36\ln n$ time.
    
    By Corollary~\ref{cor:boundONcoefficient}, at most $11 \log n$ geometric random variables will be generated, 
    and by Corollary~\ref{cor:generateGRplusEpidemic6}, with probability $\geq 1-(10/n^2)$ a given variable takes at most $36\ln n$ time to generate and propagate its maximum (total of $11 \log n\cdot36 \ln n$ time).
    
    By the union bound over all epochs, 
    the probability that generating $11 \log n +1 $ geometric random variables and propagating their maximum takes more than $(11\log n +1) \cdot 36 \ln n$ time is $\geq 1- \frac{O(\log n)}{n^2} \geq 1-1/n$ for large values of $n$.
}
\opt{full}{\proofCorTimeComplexitysix}
\opt{sub}{Finally, Corollary~\ref{cor:sumofgeometricamongAagents} bounds the \outpt\ produced by \sizest\ protocol. A proof appears in the appendix.}
\opt{full}{
}

\newpage
\section{Simulation}

\begin{figure}[ht]
	\centering
	\includegraphics[width= \textwidth, draft=false]{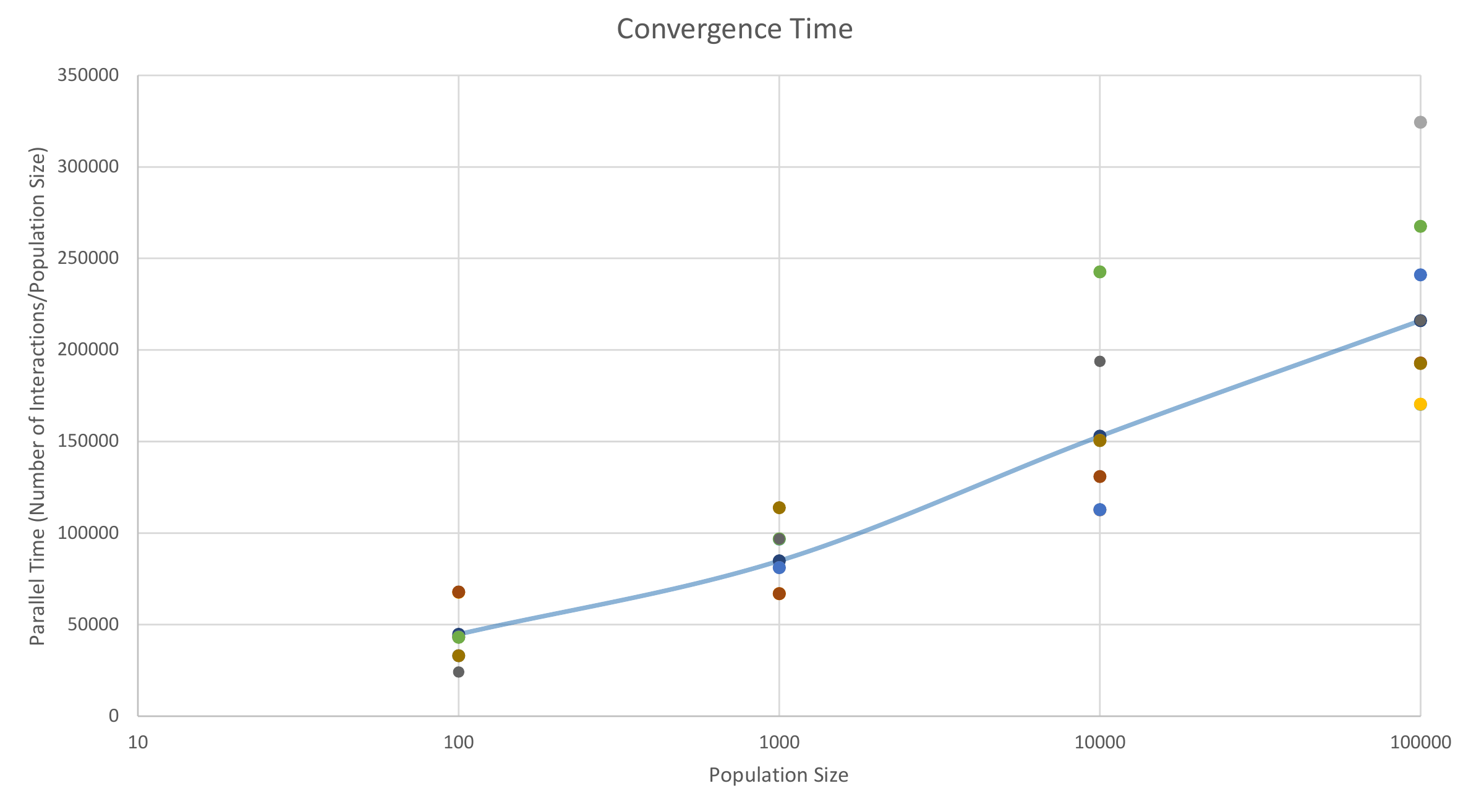}
    \caption{
        Simulated convergence time of the protocol.
        Although the proofs give only that the estimate of $\log n$ is likely to get within additive error of 5,
        in practice the estimate is always within 2,
        so this is how we define convergence in the experiment.
    	The dots indicate the convergence time of individual experiments.
    	The population size axis is logarithmic
    	(i.e., exactly $O(c \log_{10} n)$ time complexity would correspond to a straight line with slope $c$).
    	The circular dots in the plot are 10 experiments at each value of $n \in \{10^2, 10^3, 10^4, 10^5\}$. The convergence in the \sizest\ protocol happens when all agents reach $\phase = 5 \cdot \clock$.
    }
	\label{fig:time}
\end{figure}

Simulation results are shown in Fig.~\ref{fig:time}.

\newpage
\section{Chernoff bound on sums of maxima of geometric random variables}
\label{subsec:geom}

This section is aimed at proving Corollary~\ref{cor:sumofgeomConcreteK},
a Chernoff bound on the tails of $K$ independent random variables,
each of which is the maximum of $N$ independent geometric random variables with success probability $1/2$.
When applied to prove correctness of the protocol of Section~\ref{subsec:protocol},
$N$ is a value very likely to be near $n/2$,
i.e., half the population size.

    
    
    
    




\subsection{Sub-exponential random variables}

\begin{definition}\label{def:subexp}
    Let $\alpha,\beta > 0$ and let $\mathbf{X}$ be a random variable.
    We say $\mathbf{X}$ is \emph{$\alpha$-$\beta$-sub-exponential} if,
    for all $\lambda>0$,
    $\Pr{|\mathbf{X} - \E{\mathbf{X}}| \geq \lambda} \leq \alpha e^{-\lambda / \beta}$.
\end{definition}

The following lemma is well-known;
we prove it explicitly since the exact form is convenient for our purposes
but is more general than typically expressed.
It shows that exponential tail bounds for $\Pr{|\mathbf{X}-\E{\mathbf{X}}| > \lambda}$ 
give bounds on the moment-generating functions of the random variables $\mathbf{X}-\E{\mathbf{X}}$ and $\E{\mathbf{X}}-\mathbf{X}$.
The proof is modeled on Rigollet's proof of the analogous lemma for sub-gaussian random variables proven in~\cite{rigollet2015highdimensional}.


\begin{lemma}[\cite{rigollet2015highdimensional}] \label{lem:subexp-bound-mgf}
    Let $\mathbf{X}$ be a $\alpha$-$\beta$-sub-exponential random variable.
    Then for all $s \in \left[ -\frac{1}{2\beta}, \frac{1}{2\beta} \right]$, 
    we have
    $\E{ e^{ s (\mathbf{X}-\E{\mathbf{X}}) } } , \E{ e^{ s (\E{\mathbf{X}}-\mathbf{X}) } } 
    \leq 1 + 2 \alpha \beta^2 s^2$.
\end{lemma}

\begin{proof}
    Let $k \in \N^+$.
    Then
    \begin{eqnarray*}
        \E{|\mathbf{X} - \E{\mathbf{X}}|^k}
    &=&
        \int_0^\infty \Pr{|\mathbf{X} - \E{\mathbf{X}}|^k \geq \lambda} d\lambda
    =
        \int_0^\infty \Pr{|\mathbf{X} - \E{\mathbf{X}}| \geq \lambda^{1/k}} d\lambda
    \\&\leq&
        \int_0^\infty \alpha e^{- \lambda^{1/k} / \beta} d\lambda
    =
        \alpha \beta^k k \int_0^\infty e^{-u} u^{k-1} du  
        \ \ \ \ \ \ \text{substituting } u = \beta \lambda^{1/k}
    \\&=&
        \alpha \beta^k k \Gamma(k)
    =
        \alpha \beta^k k!,
    \end{eqnarray*}
    where $\Gamma(k) = \int_0^\infty e^{-u} u^{k-1} du$ 
    is the \emph{gamma function}, 
    known to equal $(k-1)!$ for $k \in \N^+$.
    Then for all $s \in \left[ -\frac{1}{2\beta}, \frac{1}{2\beta} \right]$,
    
    \begin{eqnarray*}
        \E{e^{s (\mathbf{X} - \E{\mathbf{X}}) }}
    &=&
        \E{ \sum_{k=0}^\infty 
        \frac{ ( s (\mathbf{X} - \E{\mathbf{X}}) )^k }{k!} } 
        \ \ \ \ \ \text{Taylor expansion of the exponential function}
    \\&=&
        \sum_{k=0}^\infty 
        \frac{ s^k \E{(\mathbf{X} - \E{\mathbf{X}})^k} }{k!} 
        \ \ \ \ \ \ \ \text{dominated convergence theorem}
    \\&=&
        1 + 
        s \underbrace{ \E{\mathbf{X} - \E{\mathbf{X}}} }_{= 0} + 
        \sum_{k=2}^\infty 
        \frac{ s^k \E{(\mathbf{X} - \E{\mathbf{X}})^k} }{k!}
    \\&\leq&
        1 + 
        \sum_{k=2}^\infty 
        \frac{ |s|^k \E{|\mathbf{X} - \E{\mathbf{X}}|^k} }{k!}
        \ \ \ \text{odd terms can only get larger}
    \\&\leq&
        1 + 
        \sum_{k=2}^\infty 
        \frac{ |s|^k \alpha \beta^k k! }{k!}
    =
        1 + 
        \alpha 
        \sum_{k=2}^\infty 
        \left( |s| \beta \right)^k
    =
        1 + 
        \alpha s^2 \beta^2
        \sum_{k=0}^\infty 
        \left( |s| \beta \right)^k
    \\&\leq&
        1 + \alpha \beta^2 s^2 \sum_{k=0}^\infty \frac{1}{2^k} 
        \ \ \ \ \ \ \ \ \ \ \ \text{since $|s| \leq \frac{1}{2 \beta}$}
    \\&=&
        1 + 2 \alpha \beta^2 s^2.
    \end{eqnarray*}
    The bound for $\E{ e^{ s (\E{\mathbf{X}}-\mathbf{X}) } }$
    is derived by a similar argument.
\end{proof}




The following Chernoff bound is well-known,
but stated in a more convenient form for our purposes.

\newcommand{\chernoffSubexpLemma}{Let $\alpha,\beta > 0$ and $K \in \N^+$.
Let $\mathbf{X}_1,\ldots,\mathbf{X}_K$ be i.i.d. $\alpha$-$\beta$-sub-exponential random variables.
Define $\mathbf{S} = \sum_{i=1}^K \mathbf{X}_i$.
Then for all $t \geq 0$,
\[
    \Pr{|\mathbf{S} - \E{\mathbf{S}}| \geq t } 
\leq
    2
    \frac
    { \left( 1 + \alpha / 2 \right)^K }
    { e^{t / (2 \beta)} }.
\]}

\begin{lemma}\label{lem:chernoff}
    \chernoffSubexpLemma
\end{lemma}

\begin{proof}
    Then for all $s,t>0$,
    \begin{eqnarray*}
        \Pr{\mathbf{S} - \E{\mathbf{S}} > t}
    &=& 
        \Pr{e^{ s (\mathbf{S} - \E{\mathbf{S}}) } > e^{s t}} 
    \\&\leq& 
        \frac
        { \E{e^{s (\mathbf{S} - \E{\mathbf{S}}) } } } 
        { e^{s t} } 
        \ \ \ \ \ \ \ \ \ \ \ \ \ \ \ \ \text{Markov's inequality}
    \\&=&
        e^{- s t} 
        \E{ e^{ s \left( \left( \sum_{i=1}^K \mathbf{X}_i \right) - \E{\mathbf{S}} \right) } }
    \\&=&
        e^{- s t} 
        \E{ e^{ s \sum_{i=1}^K ( \mathbf{X}_i  - \E{\mathbf{X}_i} ) } }
        \ \ \ \ \text{linearity of expectation}
    \\&=& 
        e^{- s t} 
        \E{ \prod_{i=1}^K e^{s ( \mathbf{X}_i - \E{\mathbf{X}_i} ) } } 
    \\&=& 
        e^{- s t} 
        \prod_{i=1}^K 
        \E{ e^{s ( \mathbf{X}_i - \E{\mathbf{X}_i} ) } }. 
        \ \ \ \ \text{independence of the $\mathbf{X}_i$'s}
    \end{eqnarray*}
    
    By Lemma~\ref{lem:subexp-bound-mgf},
    for all $|s| \leq \frac{1}{2\beta}$,
    $\E{ e^{s (\mathbf{X}_i - \E{\mathbf{X}_i}) } } \leq 1 + 2 \alpha \beta^2 s^2$,
    so letting $s = \frac{1}{2\beta}$, 
    \[
        \Pr{ \mathbf{S} - \E{\mathbf{S}} > t }
    \leq 
        e^{- s t}
        \left( 1 + 2 \alpha \beta^2 s^2 \right)^K
    = 
        e^{- t / (2 \beta)}
        \left( 1 + \alpha / 2 \right)^K.
    \]
    The proof that 
    $
        \Pr{ \E{\mathbf{S}} - \mathbf{S} \geq t } 
        <
        e^{- t / (2 \beta)}
        \left( 1 + \alpha / 2 \right)^K
    $
    is symmetric.
    By the union bound,
    $
        \Pr{ |\mathbf{S}-\E{\mathbf{S}}| \geq t } 
    < 
        2 \cdot e^{- t / (2 \beta)}
        \left( 1 + \alpha / 2 \right)^K.
    $
\end{proof}

\subsection{Geometric random variables and their maximum}

We say $\mathbf{G}$ is a \emph{$p$-geometric random variable}
if it is the number of consecutive flips until the first $H$ (including the $H$),
when flipping a coin with $\Pr{H}=p$.
Thus $\E{\mathbf{G}} = \frac{1}{p}$; 
in particular $\E{\mathbf{G}} = 2$ 
if $p = \frac{1}{2}$.

Defining $\mathbf{M} = \max\limits_{1 \leq i \leq N} \mathbf{G}_i$,
where each $\mathbf{G}_i$ is an i.i.d.~$\frac{1}{2}$-geometric random variable,
it is known~\cite{eisenberg2008expectation} that $\E{\mathbf{M}} \approx \log N$.
Lemma~\ref{lem:maxGeomTailBound} 
shows a tail bound on $\mathbf{M}$ for general $p$-geometric random variables,
which we will later apply to the case $p=\frac{1}{2}$.

We first require a technical lemma relating $\E{\mathbf{M}}$ and $\log N$ more precisely,
and more generally for $p$-geometric random variables for $p \neq \frac{1}{2}$.
Let $H_n = \sum_{i=1}^N \frac{1}{N}$ be the $N$'th harmonic number.
Let $\gamma = \lim\limits_{N\to\infty} (H_n - \ln N) \approx 0.577$
be the Euler-Mascheroni constant; 
for all $N \geq 50$ we have $H_n - \ln N -\gamma \leq 0.01$.

\begin{lemma}\label{lem:known_bound_for_expected_value_of_max}
    Let $\mathbf{G}_1,\ldots,\mathbf{G}_N$ be i.i.d.~$p$-geometric random variables
    with $q = 1-p \geq \frac{1}{e}$,
    $N \geq 50$,
    and
    let $\mathbf{M} = \max\limits_{1 \leq i \leq N} \mathbf{G}_i$.
    Let $\epsilon_1 = 0.01$ and $\epsilon_2 = 0.0006$.
    Then for all $\lambda > 0$, 
    $\frac{\ln(N)+\gamma}{\ln 1/q}+1/2 - \epsilon_2 < \E{\mathbf{M}} < \frac{\ln(N) + \gamma +\epsilon_1}{\ln 1/q}+ 1/2 + \epsilon_2$; particularly for $q=p=1/2$, we have:
    $\log N +1 < \E{\mathbf{M}}<\log N +3/2$.
\end{lemma}

\begin{proof}
    Eisenberg~\cite{eisenberg2008expectation} showed that 
    if $q \geq \frac{1}{e}$,
    then
    $\frac{1}{\lambda} H_n - 0.0006 \leq \E{\mathbf{M}}-1/2 < \frac{1}{\lambda} H_n+0.0006$, 
    where $q = e^{-\lambda}$, i.e. $\lambda = \ln(1/q)$.
    Thus
    $ \frac{1}{\lambda} H_n +1/2 - \epsilon_2 \leq \E{\mathbf{M}} < \frac{1}{\lambda} H_n + 1/2 + \epsilon_2$ 
    i.e.,
    $\frac{\ln N+\gamma}{\ln 1/q}+1/2 - \epsilon_2 < \E{\mathbf{M}} < \frac{\ln N + \gamma +\epsilon_1}{\ln 1/q}+ 1/2 + \epsilon_2$.
\end{proof}

\begin{lemma} \label{lem:maxGeomTailBound}
    Let $\mathbf{G}_1,\ldots,\mathbf{G}_N$ be i.i.d.~$p$-geometric random variables
    with $q = 1-p \geq \frac{1}{e}$,
    $N \geq 50$,
    and
    let $\mathbf{M} = \max\limits_{1 \leq i \leq N} \mathbf{G}_i$.
    Let $\epsilon_1 = 0.01$ and $\epsilon_2 = 0.0006$.
    Then for all $\lambda > 0$, 
    $
        \Pr{ \E{\mathbf{M}}-\mathbf{M} \geq \lambda } 
    \leq 
        \exp{ - q^{1/2 + \epsilon_2 - (\gamma+\epsilon_1)\ln q - \lambda} }
    $
    and
    $
        \Pr{ \mathbf{M}-\E{\mathbf{M}} \geq \lambda } 
    \leq 
        q^{\lambda- 1/2 - \epsilon_2- \gamma \ln q} + q^{2\lambda- 1 - 2\epsilon_2-2 \gamma \ln q}.
    $
\end{lemma}

\begin{proof}
    For each $t \in \N$, 
    $
        \Pr{\mathbf{G}_i \geq t}
        = q^{t-1}
    $,
    so
    $
        \Pr{\mathbf{G}_i \leq t}
        = 1 - \Pr{\mathbf{G}_i \geq t+1}
        = 1 - q^{t}.
    $
    
    Since the $\mathbf{G}_i$'s are independent,
    $
        \Pr{\mathbf{M} \leq t} 
        = \prod_{i=1}^N \left( 1 - q^{t} \right)
        = \left( 1 - q^{t} \right)^N.
    $
    
    
    
    Below we use Lemma~\ref{lem:known_bound_for_expected_value_of_max} and the inequalities
    $e^x \left( 1 - \frac{x^2}{N} \right) 
    \leq \left( 1+\frac{x}{N} \right)^N 
    \leq e^x$
    for $N > 1, |x| < N$.

    Setting $t = \E{\mathbf{M}} - \lambda$, we have
    \begin{eqnarray*}
        \Pr{\mathbf{M} \leq \E{\mathbf{M}} - \lambda}
    &=&
        \left( 1 - q^{t} \right)^N
    \\&=&
        \left( 1 - q^{(\E{\mathbf{M}} - \lambda)} \right)^N
    \\&<&
        \left( 1 - q^{\log_{1/q} N - (\gamma+\epsilon_1)\ln q + 1/2 + \epsilon_2 - \lambda} \right)^N
    \\&=&
        \left( 1 - q^{\log_{1/q} N} q^{1/2 + \epsilon_2 - (\gamma+\epsilon_1)\ln q - \lambda} \right)^N
    \\&=&
        \left( 1 - \frac{ q^{1/2 + \epsilon_2 - (\gamma+\epsilon_1)\ln q - \lambda} } {N} \right)^N
    \\&<&
        \exp{ - q^{1/2 + \epsilon_2 - (\gamma+\epsilon_1)\ln q - \lambda} }
    \end{eqnarray*}
    The last inequality is true since 
        $\left( 1+\frac{x}{N} \right)^N \leq e^x$.
        
    Similarly,
    letting $t = \E{\mathbf{M}} + \lambda - 1$, we have
    \begin{eqnarray*}
    &&
        \Pr{\mathbf{M} \geq \E{\mathbf{M}} + \lambda}
    \\&=&
        1 - \Pr{\mathbf{M} \leq \E{\mathbf{M}} + \lambda - 1}
    \\&=&
        1 - \left( 1 - q^{t} \right)^N
    \\&=&
        1 - \left( 1 - q^{\E{\mathbf{M}} + \lambda -1} \right)^N
    \\&<&
        1 - \left( 1 - q^{\log_{1/q} N + 1/2 - \epsilon_2 - \gamma \ln q+ \lambda -1} \right)^N
    \\&=&
        1 - \left( 1 - q^{\log_{1/q} N} q^{\lambda - 1/2 - \epsilon_2 - \gamma \ln q} \right)^N
    \\&=&
        1 - \left( 1 - \frac{q^{\lambda - 1/2 - \epsilon_2 - \gamma \ln q}} {N} \right)^N
    \\&<&
        1 - 
        \exp{ - q^{\lambda - 1/2 - \epsilon_2 - \gamma \ln q} }
        \left( 1 - \frac{ q^{2(\lambda - 1/2 - \epsilon_2 - \gamma \ln q)} }{N} \right)
        \ \ \ \ \text{since } 
        e^x \left( 1 - \frac{x^2}{N} \right) 
        \leq \left( 1+\frac{x}{N} \right)^N
    \\&=&
        1 - 
        \exp{ - q^{\lambda- 1/2 - \epsilon_2 - \gamma \ln q} + \ln \left( 1 - \frac{q^{2\lambda- 1 - 2\epsilon_2  - 2\gamma \ln q}}{N} \right)}
    \\&\leq&
        q^{\lambda- 1/2 - \epsilon_2 - \gamma \ln q} - \ln \left( 1 - \frac{q^{2\lambda- 1 - 2\epsilon_2 -2 \gamma \ln q}}{N} \right)
        \ \ \ \ \text{since } 1 - e^x \leq - x
    \\&\leq&
        q^{\lambda- 1/2 - \epsilon_2- \gamma \ln q} + \frac{2 q^{2\lambda- 1 - 2\epsilon_2-2 \gamma \ln q}}{N}
        \ \ \ \ \text{since } \ln(1-x) \geq -2x \text{ if } x < 0.7
    \\&\leq&
        q^{\lambda- 1/2 - \epsilon_2- \gamma \ln q} + q^{2\lambda- 1 - 2\epsilon_2-2 \gamma \ln q}
        \ \ \ \ \text{since } N \geq 2.   \qedhere
    \end{eqnarray*}
\end{proof}

The following corollary for the special case of $p=\frac{1}{2}$ is used for our main result,
showing that a maximum of $\frac{1}{2}$-geometric random variables is 
$\alpha$-$\beta$-sub-exponential for 
$\alpha=3.31, \beta=2$.

\begin{corollary} \label{cor:half-geom-is-sub-exp}
    Let $\mathbf{G}_1,\ldots,\mathbf{G}_N$ be i.i.d. $\frac{1}{2}$-geometric random variables,
    $N \geq 50$,
    and let $\mathbf{M} = \max\limits_{1 \leq i \leq N} \mathbf{G}_i$.
    Then for all $\lambda > 0$,
    $\Pr{ |\mathbf{M}-\E{\mathbf{M}}| \geq \lambda } < 3.31 e^{- \lambda / 2}$.
\end{corollary}

\begin{proof}
    By Lemma~\ref{lem:maxGeomTailBound} and the union bound,
    \begin{eqnarray*}
        \Pr{|\mathbf{M} - \E{\mathbf{M}}| \geq \lambda}
    &<&
        \exp{ - q^{1/2 +\epsilon_2 - (\gamma+\epsilon_1)\ln q - \lambda} }
        + q^{\lambda- 1/2 - \epsilon_2- \gamma \ln q} 
        + q^{2\lambda- 1 - 2\epsilon_2-2 \gamma \ln q}
    \\&=&
        \exp{ - 2^{\lambda - (\gamma+\epsilon_1)\ln 2 - 1/2 - \epsilon_2} }
        + 2^{1/2 + \epsilon_2-\lambda-\gamma \ln 2} 
        + 2^{1 + 2\epsilon_2-2\lambda-2\gamma \ln 2}
    \\&<&
        3.31 e^{-\lambda / 2}.
        \ \ \ \ \text{justified below}
    \end{eqnarray*}
    To see the final inequality,
    note that 
    
    \begin{eqnarray*}
        \exp{ - 2^{\lambda - (\gamma+\epsilon_1)\ln 2 -1/2 -\epsilon_2} } 
    &<&
        \exp{ - 2^{\lambda - 1} } 
    \\&=&
        \exp{ - 2^{\lambda}/2 } 
    \\&\leq&
        \exp{ - \lambda/2}
    \end{eqnarray*}

    \begin{eqnarray*}
        2^{1/2 +\epsilon_2-\lambda-\gamma \ln 2} 
    &=&
        2^{1/2 +\epsilon_2-\gamma \ln 2} \cdot 2^{-\lambda}
    \\&=&
        2^{1/2 +\epsilon_2-\gamma \ln 2} \cdot 4^{-\lambda/2}
    \\&<&
        2^{1/2 +\epsilon_2-\gamma \ln 2} \cdot e^{-\lambda/2}
    \\&<&
        1.1 \cdot e^{-\lambda/2}.
    \end{eqnarray*}

    \begin{eqnarray*}
        2^{1 + 2\epsilon_2-2\lambda-2\gamma \ln 2}
    &=&
        2^{1 + 2\epsilon_2-2\gamma \ln 2} \cdot 2^{-2\lambda}
    \\&=&
        2^{1 + 2\epsilon_2-2\gamma \ln 2} \cdot 16^{-\lambda/2}
    \\&<&
        (1.1)^2 \cdot e^{-\lambda/2}.
    \end{eqnarray*}
    So, their sum is less than  $3.31 e^{-\lambda/2}$. 
\end{proof}

The following lemma bounds the maximum of $N$ $\frac{1}{2}$-geometric random variables for the special cases of one lower and one upper threshold,
which is stronger than the bounds given by Corollary~\ref{cor:half-geom-is-sub-exp}.

\begin{lemma}
\label{cor:half-geom-tighter-bounds}
    Let $\mathbf{G}_1,\ldots,\mathbf{G}_N$ be i.i.d.~$\frac{1}{2}$-geometric random variables, $N\geq50$, 
    and $\mathbf{M} = \max\limits_{1 \leq i \leq N} \mathbf{G}_i$.
    Then 
    $\Pr{ \mathbf{M} \geq 2 \log N } < N^{-1}$
    and
    $\Pr{ \mathbf{M} \leq \log N - \log \ln N } < N^{-1}$.
\end{lemma}

\begin{proof}
     For any $i$,
    $\Pr{\mathbf{G}_i \geq \log N- \log \ln N} 
    = \left( \frac{1}{2} \right)^{(\log N -\log \ln N)} 
    = \frac{\ln N}{N}$.
   Since the $\mathbf{G}_i$'s are independent,
    $
        \Pr{\mathbf{M} < \log N - \log \ln N}
    =
        \Pr{(\forall i)\ \mathbf{G}_i < \log N - \log \ln N}
    =
        \left( 1-\frac{\ln N}{N} \right)^N
    \leq
        e^{-{\ln N}} 
    =
        N^{-1}.
    $
    For the upper bound, for any $i$,
    $\Pr{\mathbf{G}_i \geq 2\log N} 
    = \left( \frac{1}{2} \right)^{2\log N} 
    = N^{-2}$. 
    By the union bound, 
    $\Pr{\mathbf{M} \geq 2\log N}
    = \Pr{(\exists i)\ \mathbf{G}_i \geq 2\log N} 
    \leq N^{-1}$.
\end{proof}

\begin{lemma}\label{lem:chernoff-geometric}
    Let $N,K \in \N^+$, $N\geq50$.
    Let $\mathbf{M}_1,\ldots,\mathbf{M}_K$ be i.i.d. random variables,
    each of which is the maximum of $N$ i.i.d. $\frac{1}{2}$-geometric random variables.
    Define $\mathbf{S} = \sum_{i=1}^K \mathbf{M}_i$.
    Then for all $t \geq 0$,
    $\Pr{|\mathbf{S} - \E{\mathbf{S}}| \geq t } 
    \leq 2 \cdot  e^{K - t/4}.$
\end{lemma}

\begin{proof}
    %
    By Corollary~\ref{cor:half-geom-is-sub-exp} 
    and Lemma~\ref{lem:chernoff},
    for $\alpha = 3.31 < 2e-2$ and $\beta = 2$,
    we have
    \begin{eqnarray*}
        \Pr{|\mathbf{S} - \E{\mathbf{S}}| \geq t } 
    &<&
        2
        \frac
        { \left( 1 + \alpha / 2 \right)^K }
        { e^{t / (2 \beta)} }
    <
        2 
        \frac
        { \left( 1+ \frac{2e-2}{2} \right)^K }
        { e^{t / 4} } 
    =
        2 \cdot  (e)^K \cdot  e^{-t/4}
    =
        2 \cdot  e^{K - t/4}. \qedhere
    \end{eqnarray*}
\end{proof}

\begin{corollary}\label{cor:sumofgeom}
    Let 
    $a > 4$,
    $N \in \N^+$, $N>50$,  
    $K \geq \frac{\ln N}{\frac{a}{4} - 1}$, and $\delta_0 = 1/2+\gamma/\ln 2 - \epsilon_2$. 
    Let $\mathbf{M}_1,\ldots,\mathbf{M}_K$ be i.i.d. random variables,
    each of which is the maximum of $N$ i.i.d. $\frac{1}{2}$-geometric random variables.
    Define $\mathbf{S} = \sum_{i=1}^K \mathbf{M}_i$.
    Then
    $$
        \Pr{ \left| \frac{\mathbf{S}}{K} - \log N - \delta_0 \right| \geq a } 
    \leq 
        \frac{2}{N}.
    $$
\end{corollary}

\begin{proof}
    We first manipulate the expression in the conclusion of the corollary
    to put it in a form where we can apply Lemma~\ref{lem:chernoff-geometric}.
            
    \begin{eqnarray*}
    &&
        \Pr{ \frac{\mathbf{S}}{K} - \log N -\delta_0 \geq a } 
    \\&=&
        \Pr{ \mathbf{S} - K \left(\log N +1/2+ \gamma/\ln 2-\epsilon_2 \right)  \geq a K} 
    \\&<&
        \Pr{ \mathbf{S} - \E{\mathbf{S}}  \geq 
        aK }.
    \end{eqnarray*}
    Since $K \left( \log N + 1/2 +\gamma/\ln 2 - \epsilon_2 \right) \leq E[\mathbf{S}]$.
    
    \begin{eqnarray*}
    &&
        \Pr{ \log N + \delta_0 - \frac{\mathbf{S}}{K} \geq a } 
    \\&=&
        \Pr{ K \left ( \log N +1/2 +\gamma/\ln 2 - \epsilon_2 \right)- \mathbf{S} \geq aK }
    \\&=&
        \Pr{ K \left(\log N + 1/2 +\gamma/\ln 2 + \epsilon_1/\ln 2 + \epsilon_2 \right) - \mathbf{S} \geq aK+ \left( \epsilon_1/\ln 2 + 2\epsilon_2 \right)K } 
    \\&<&
        \Pr{ \E{\mathbf{S}} - \mathbf{S} \geq (a+ \epsilon_1/\ln 2 +2\epsilon_2)K}
    \\&<&
        \Pr{ \E{\mathbf{S}} - \mathbf{S} \geq aK }. 
    \end{eqnarray*}
    Because $ E[\mathbf{S}] < K \left( \log N + 1/2 +\frac{\gamma + \epsilon_1}{\ln 2}+ \epsilon_2 \right) $.
    
    Since the events 
    $
        \frac{\mathbf{S}}{K} - \log N - \delta_0 \geq a
    $
    and
    $
        \log N + \delta_0 - \frac{\mathbf{S}}{K} \geq a
    $
    are disjoint,
    and the events
    $
        \mathbf{S} - \E{\mathbf{S}}  \geq aK
    $
    and
    $
        \E{\mathbf{S}} - \mathbf{S}  \geq aK
    $
    are disjoint,
    the union bound holds with equality,
    so
    \begin{eqnarray*}
        \Pr{ \left| \frac{\mathbf{S}}{K} - \log N - \delta_0 \right| \geq a } 
    &=&
        \Pr{ \frac{\mathbf{S}}{K} - \log N -\delta_0 \geq a } 
    +
        \Pr{ \log N + \delta_0 - \frac{\mathbf{S}}{K} \geq a }
    \\&<&
        \Pr{ \mathbf{S} - \E{\mathbf{S}}  \geq aK }
    +
        \Pr{ \E{\mathbf{S}} - \mathbf{S}  \geq aK }
    \\&=&
        \Pr{ | \mathbf{S} - \E{\mathbf{S}} |  \geq aK }.
    \end{eqnarray*}


    Let $t = aK$.
    Applying Lemma~\ref{lem:chernoff-geometric} with these values of $K$ and $t$,
    \begin{eqnarray*}
        \Pr{ \left| \frac{\mathbf{S}}{K} - \log N - \delta_0 \right| \geq a } 
    &<&
        \Pr{ |\mathbf{S} - \E{\mathbf{S}}|  \geq aK }     
    \\&=&
        \Pr{ \left| \mathbf{S} - \E{\mathbf{S}} \right| \geq t } 
    \\&\leq&
        2 \cdot  e^{K - t/4}
    \\&=&
        2 \cdot  e^{K (1 - \frac{a}{4})}
    \\&=&
        2 \cdot  e^{- K (\frac{a}{4} - 1)}
    \\&\leq&
        2 \cdot  e^{- \frac{\ln N}{(\frac{a}{4} - 1)} (\frac{a}{4} - 1) }
    \\&=& 
        2 \cdot  e^{- \ln N}
    \\&=& 
        \frac{2}{N}. \qedhere
    \end{eqnarray*}
\end{proof}

For example,
choosing $a = \ln 2 + 4 < \logSizeError$ means we can choose
$K 
\geq \frac{\ln N}{\frac{a}{4} - 1} 
= \frac{\ln N}{(\ln(2) + 4)/4 - 1} 
= \frac{\ln N}{\ln(2) / 4}
= 4\log_2 N$:

\begin{corollary}\label{cor:sumofgeomConcreteK}
    Let 
    $N \in \N^+$, $N\geq50$, 
    $K \geq 4 \log N$. 
    Let $\mathbf{M}_1,\ldots,\mathbf{M}_K$ be i.i.d. random variables,
    each of which is the maximum of $N$ i.i.d. $\frac{1}{2}$-geometric random variables.
    Define $\mathbf{S} = \sum_{i=1}^K \mathbf{M}_i$.
    Then
    $$
        \Pr{ \left| \frac{\mathbf{S}}{K} - \log N \right| \geq \logSizeError } 
    \leq 
        \frac{2}{N}.
    $$
\end{corollary}

\newpage

\section{Timer lemma} \label{app:termination}

\begin{lemma}\label{lem:balls-bins-decay}
    Let $0 < \delta \leq \frac{1}{2}$.
    Let $0 < k \leq n$ and $m$ be positive integers.
    Suppose we have $n$ bins, of which $k$ are initially empty, 
    and we throw $m$ additional balls randomly into the $n$ bins.
    Then 
    $
        \Pr{\leq \delta k \text{ bins remain empty}} 
    < 
        ( 2 \delta e^{m/n} )^{\delta k}.
    $
\end{lemma}

\begin{proof}
    When there are $i$ bins empty, the probability that the next ball fills an empty bin is $\frac{i}{n}$.
    Thus, the number of balls needed until $\leq \delta k$ bins are empty is a sum 
    $\mathbf{S} = \sum_{i=\delta k+1}^{k} \mathbf{G}_i$ 
    of independent geometric random variables 
    $\mathbf{G}_{\delta k+1},\ldots,\mathbf{G}_{k}$,
    where $\mathbf{G}_i$ has $p_i = \Pr{\text{success}} = \frac{i}{n}$,
    ``success'' representing the event of throwing a ball into one of the $k$ initially empty bins.
    
    The moment-generating function of a geometric random variable $\mathbf{G}$ with $\Pr{\text{success}} = p$,
    defined whenever $\theta < - \ln (1-p)$~\cite{mitzenmacher2005probability},
    is 
    \[
        \E{ e^{ \theta \mathbf{G} } }
    = 
        \frac
        { p e^\theta }
        { 1 - (1-p) e^\theta }
    = 
        \frac
        { p }
        { e^{-\theta} - 1 + p}
    \leq
        \frac
        { p }
        { p - \theta },
    \]
    where the last inequality follows from 
    $e^{x} - 1 \geq x$ for all $x \in \R$.
    Thus for each $i \in \{\delta k, \ldots,k\}$,
    \[
        \E{ e^{ \theta \mathbf{G}_i } }
    \leq
        \frac
        { \frac{i}{n} }
        { \frac{i}{n} - \theta }
    =
        \frac
        { i }
        { i - \theta n }.
    \]
    
    By independence of the $\mathbf{G}_i$'s,
    the moment-generating function of the sum $\mathbf{S}$ is
    \[ 
        \E{ e^{ \theta \mathbf{S} } }
    = 
        \E{ e^{ \theta \sum_{i=\delta k+1}^{k} \mathbf{G_{i}} } }
    = 
        \prod_{i=\delta k+1}^{k} 
        \E{ e^{ \theta \mathbf{G_{i}} } }
    \leq 
        \prod_{i=\delta k+1}^{k} 
        \frac
        { i }
        { i - \theta n }.
    \] 
    Setting $\theta = - \frac{\delta k}{n}$,
    and using the fact that $\delta \leq \frac{1}{2}$ to cancel terms,
    we have
    \begin{eqnarray*}
    &&
        \E{ e^{ \theta \mathbf{S} } }
    \\&\leq&
        \prod_{i=\delta k+1}^{k} 
        \frac
        { i }
        { i + \delta k }
    =
        \left(
            \frac 
            { \delta k + 1 } 
            { \delta k + 1 + \delta k }
        \right)
        \left(
            \frac 
            { \delta k + 2 } 
            { \delta k + 2 + \delta k }
        \right)
        \ldots
        \left(
            \frac 
            { k } 
            { k + \delta k }
        \right)
    \\&=&
        \frac
        { (\delta k + 1) \ldots (\delta k + \delta k) }
        { 1 }
        \cdot
        \frac 
        { (\delta k + 1 + \delta k) \ldots (k) } 
        { (\delta k + 1 + \delta k) \ldots (k) }
        \cdot
        \frac 
        { 1 } 
        { (k + 1) \ldots (k + \delta k) }
    \\&=&
        \frac 
        { (\delta k + 1) \ldots (\delta k + \delta k) } 
        { (k + 1) \ldots (k + \delta k) }
    <
        \left(
            \frac 
            { 2 \delta k } 
            { k }
        \right)^{\delta k}
    =
        (2 \delta)^{\delta k}.
    \end{eqnarray*}

    The event that 
    throwing $m$ balls results in at most $\delta k$ empty bins 
    is equivalent to the event that 
    $\mathbf{S} \leq m$.
    By Markov's inequality,
    since $\theta = - \frac{\delta k}{n} < 0$,
    \[
        \Pr{\mathbf{S} \leq m}
    =
        \Pr{e^{\theta \mathbf{S}} \geq e^{\theta m}}
    \leq
        \frac
        { \E{ e^{\theta S} } }
        { e^{\theta m} }
    <
        (2 \delta)^{\delta k}
        e^{\frac{\delta k}{n} m}
    =
        ( 2 \delta e^{m/n} )^{\delta k}. \qedhere
    \]
\end{proof}

We say a transition 
\emph{consumes} a state $s$ if executing the transition strictly reduces the count of $s$,
and that the transition \emph{produces} $s$ if it strictly increases the count of $s$.
The next lemma bounds the rate of consumption of $s$,
showing that the count of $s$ cannot decrease too quickly.
It also makes the observation that,
since we are reasoning about $s$ assuming that it is only consumed,
we can upper-bound the probability of the count of $s$ dropping below $\delta k$ 
at \emph{any} time $t \in [0,T]$,
not just at time $t=T$.

\begin{lemma} \label{lem:chernoff-decay}
    Let $s$ be a state in a population protocol,
    let $0 < \delta \leq \frac{1}{2}$,
    and let $k$ be the count of $s$ at time 0.
    Let $\timeRand{t}{s}$ denote the count of $s$ at time $t$.
    Then for all $T > 0$,
    \[
        \Pr{ (\exists t \in [0,T])\ \timeRand{t}{s} \leq \delta k }
    \leq
        ( 2 \delta e^{3T} )^{\delta k}.
    \]
\end{lemma}

\begin{proof}
    $s$ may be produced and consumed.
    To establish that the count of $s$ remains large for a constant amount of time, 
    in the worst case we assume that $s$ is only consumed.
    We also make the worst case assumption that each time an agent in state $s$ is picked for a transition,
    it changes state and we consume that copy of $s$.
    We further make the worst-case assumption that if both agents are in state $s$,
    both change to a different state.
    
    The following \emph{almost} works:
    model each transition as throwing \emph{two} balls into bins,
    where each agent is a bin,
    considered ``empty'' if it is in state $s$.
    Each time a transition picks an agent,
    this puts a ball into the bin that agent represents.
    Thus, the number of balls in a bin represents the total number of times that the agent interacts.
    However, these are not identically distributed processes,
    since a bin may be picked twice consecutively in the balls-and-bins distribution,
    whereas when agents are picked two at a time,
    the two agents are guaranteed to be unequal.
    Thus the actual distribution has slightly higher probability of fewer empty bins than the simplified ``throw-two-balls-every-transition'' approximation.
    
    So instead, consider the distribution of empty bins given by throwing \emph{three} balls for every transition.
    Suppose $p \in \{\delta k+1,\ldots,k\}$ bins out of $n$ are currently empty.
    After the next three balls,
    the number of empty bins $\mathbf{E}_3$ will be $p$, $p-1$, $p-2$, or $p-3$.
    We have that 
    \begin{eqnarray*}
        \Pr{\mathbf{E}_3=p} 
    &=& 
        \left( \frac{n-p}{n} \right)^3,
    \\
        \Pr{\mathbf{E}_3=p-1}
    &=& 
        \left( \frac{n-p}{n} \right)^2 
        \cdot
        \frac{p}{n}
    \\&& +
        \frac{n-p}{n}
        \cdot
        \frac{p}{n}
        \cdot
        \frac{n-p-1}{n}
    \\&& +
        \frac{p}{n}
        \cdot
        \left( \frac{n-p-1}{n} \right)^2,
    \\
        \Pr{\mathbf{E}_3=p-2}
    &=& 
        \frac{n-p}{n}
        \cdot
        \frac{p}{n}
        \cdot
        \frac{p-1}{n}
    \\&& +
        \frac{p}{n}
        \cdot
        \frac{n-p-1}{n}
        \cdot
        \frac{p-1}{n}
    \\&& +
        \frac{p}{n}
        \cdot
        \frac{p-1}{n}
        \cdot
        \frac{n-p-2}{n},
    \\
        \Pr{\mathbf{E}_3=p-3}
    &=&
        \frac{p}{n}
        \cdot
        \frac{p-1}{n}
        \cdot
        \frac{p-2}{n}.
    \end{eqnarray*}
    Compare this to the true distribution $\mathbf{E}_2$ of the number of empty bins after one interaction,
    where two unequal bins are picked at random each to get a ball.
    Then
    \begin{eqnarray*}
        \Pr{\mathbf{E}_2=p} 
    &=& 
        \frac{ \binom{n-p}{2} }{ \binom{n}{2} }
    = 
        \frac{ (n-p) (n-p-1) }{ n(n-1) },
    \\
        \Pr{\mathbf{E}_2=p-1}
    &=& 
        \frac{ n-p }{ n }
        \cdot
        \frac{ p }{ n },
    \\
        \Pr{\mathbf{E}_2=p-2}
    &=& 
        \frac{ \binom{p}{2} }{ \binom{n}{2} }
        \cdot
        \frac{ \binom{p-1}{2} }{ \binom{n}{2} }
    =
        \frac{ p(p-1) }{ n(n-1) }
        \cdot
        \frac{ (p-1)(p-2) }{ n(n-1) }
    \\
        \Pr{\mathbf{E}_2=p-3}
    &=&
        0
    \end{eqnarray*}
    It can be verified by inspection that for each $\ell \in \{p-3,p-2,p-1,p\}$,
    \begin{eqnarray*}
        \Pr{\mathbf{E}_2 \leq \ell} 
    &=&
        \sum_{\ell' \in \{p-3,\ldots,\ell\}} \Pr{\mathbf{E}_2 = \ell'}
    \\&<& 
        \sum_{\ell' \in \{p-3,\ldots,\ell\}} \Pr{\mathbf{E}_3 = \ell'}
    \\&=&
        \Pr{\mathbf{E}_3 \leq \ell}.
    \end{eqnarray*}
    Thus,
    the distribution of empty bins given by throwing three balls independently at random stochastically dominates the true distribution of empty bins after one interaction.
    
    The number of interactions in time $T$ is $Tn$.
    Using the stochastically dominating distribution above,
    we model this as throwing $m = 3Tn$ balls independently.
    By Lemma~\ref{lem:balls-bins-decay},
    \[
        \Pr{ (\exists t \in [0,T])\ \timeRand{t}{s} \leq \delta k }
    \leq
        ( 2 \delta e^{m/n} )^{\delta k}
    =
        ( 2 \delta e^{3T} )^{\delta k}. \qedhere
    \]
\end{proof}

The following corollary with $\delta = \frac{1}{81}$ and $T=1$ 
states that within time $1$,
it is unlikely for the count of any state to decrease by more than factor $81$ 
from $k$ to $k / 81$.

\begin{corollary} \label{cor:chernoff-decay}
    Let $s$ be a state in a population protocol,
    and let $k$ be the count of $s$ at time 0.
    Let $\timeRand{t}{s}$ denote the count of $s$ at time $t$.
    Then
    \[
        \Pr{ (\exists t \in [0,1])\ \timeRand{t}{s} \leq k / 81 }
    \leq
        2^{- k / 81}.
    \]
\end{corollary}

\begin{proof}
    Note that $2 e^3 < 40.2$, 
    so setting $\delta = \frac{1}{81}$ and $T=1$ implies that
    $2 \delta e^{3T} < \frac{1}{2}$.
    Applying Lemma~\ref{lem:chernoff-decay} with 
    $\delta = \frac{1}{81}$
    and
    $T=1$,
    we have
    \[
        \Pr{ (\exists t \in [0,1])\ \timeRand{t}{s} \leq k / 81 }
    <
        ( 2 \delta e^{3T} )^{\delta k}
    <
        2^{- k / 81}. \qedhere
    \]
\end{proof}

Recall the timer lemma used in Section~\ref{sec:termination}.
The proof follows the same structure as the main theorem of~\cite{Do14},
but uses the discrete-time model of population protocols rather than the continuous-time model of chemical reaction networks.
Additionally,
care must be taken to show that although the number of states is infinite
(so clearly only a finite number can appear in finite time),
those states producible via a constant number of transitions,
whose probabilities are bounded below by a positive constant,
in sufficiently large dense configurations
are all produced in large quantity in constant time.

\restateLemTimer
\proofLemTimer

\newpage
\bibliographystyle{plain}
\bibliography{My-Library}

\end{document}